\documentclass{aa}

\usepackage{graphicx}
\usepackage{txfonts}
\usepackage[hidelinks]{hyperref}
\usepackage{xspace}
\usepackage[separate-uncertainty = true, multi-part-units=single]{siunitx}
\usepackage{tabularx}
\usepackage{booktabs}
\usepackage{multirow}
\usepackage{ragged2e}
\usepackage{orcidlink}

\newcommand{\pltscale}{0.7}
\newcommand{\spc}[2]{#1\,\textsc{\romannumeral #2}\xspace}
\newcommand{\Multi}{\texttt{MULTI}\xspace}
\newcommand{\MultiD}{\texttt{MULTI3D}\xspace}
\newcommand{\Dispatch}{\texttt{DISPATCH}\xspace}

\newcommand{\Mdis}{\texttt{M3DIS}\xspace}
\newcommand{\Turb}{\texttt{TurbospectrumNLTE}\xspace}
\newcommand{\Code}{\texttt{OpenMULTI3D}\xspace}
\newcommand{\Stagger}{\texttt{STAGGER}\xspace}
\newcommand{\Cobold}{\texttt{CO5BOLD}\xspace}
\newcommand{\Marcs}{\texttt{MARCS}\xspace}
\newcommand{\Muram}{\texttt{MURAM}\xspace}

\newcommand{\Balder}{\texttt{BALDER}\xspace}
\newcommand{\Teff}{$T_{\text{eff}}$\xspace}
\newcommand{\stellarparams}[3]{$T_{\text{eff}}=\SI{#1}{K}$, $\log g=#2$, $\mathrm{[Fe/H]}=#3$}
\DeclareSIUnit\angstrom{\text{Å}}

\begin{document}

    \title{The OpenMULTI3D stellar analysis package}

   \subtitle{A 3D non-LTE spectrum synthesis code for AFGKM-type stars
   \\ within the DISPATCH framework}

   \author{
    Richard Hoppe$^{1,2}$\orcidlink{0000-0002-8451-6260},
    Maria Bergemann$^{1}$\orcidlink{0000-0002-9908-5571},
    Philipp Eitner$^{1}$\orcidlink{0009-0007-4502-8081},
    \AA{}ke Nordlund$^{3}$\orcidlink{0000-0002-2219-0541},
    Jorrit Leenaarts$^{4}$\orcidlink{0000-0003-4936-4211},
    Andrius Popovas$^{5,6}$\orcidlink{0000-0003-4436-0764}\,and
    Nicholas Storm$^{1,2}$\orcidlink{0000-0002-5259-3974}
    }

    \institute{
    $^{1}$ Max-Planck-Institut für Astronomie, 69117 Heidelberg, Germany\\
    $^{2}$ Ruprecht-Karls Universität, 69117 Heidelberg, Germany\\
    $^{3}$ University of Copenhagen, Niels Bohr Institute, Copenhagen, DK-2100, Denmark \\
    $^{4}$ Institute for Solar Physics, Department of Astronomy, Stockholm University, AlbaNova University Centre, SE-106 91 Stockholm, Sweden \\
    $^{5}$ Rosseland Centre for Solar Physics, University of Oslo, P.O. Box 1029, Blindern, NO-0315 Oslo, Norway \\
    $^{6}$ Institute of Theoretical Astrophysics, University of Oslo, P.O. Box 1029, Blindern, NO-0315 Oslo, Norway \\
    \email{hoppe@mpia.de}
    }

   \date{Received XX; accepted XX}

  \abstract
   {Upcoming large spectroscopic surveys such as WEAVE, 4MOST, and SDSS-V will provide millions of high-resolution stellar spectra. This motivates the development of efficient codes to perform three-dimensional (3D) non-local thermodynamic equilibrium (non-LTE) radiative transfer (RT) calculations, in order to meet the requirements on efficiency and data analysis quality set by these surveys.}
   {We present a new open access 3D non-LTE spectrum synthesis code \Code, based on its predecessor \MultiD, which was never released. This code can be used to compute large grids of spectra of AFGKM-type stars and will aid the exploitation of large-scale spectroscopic surveys such as 4MOST.}
   {We summarise the main input micro- and macro-physics of the code, describe its numerical implementation and provide selected test cases.}
   {The outputs of the code are compared with standard results in the literature, including the NLTE departure coefficients in 1D and 3D, the 3D NLTE abundance corrections, 3D synthetic spectral energy distributions for three types of stars, and 3D NLTE Balmer-line profiles of a metal-poor red giant. We also demonstrate how the code can be used to compute time-dependent spectra of stars, which may be relevant in precision stellar physics and analyses of exoplanet spectra.}
   {\Code offers an accessible and efficient platform for high-fidelity stellar analysis in the era of large-scale spectroscopic surveys.}


\titlerunning{OpenMULTI3D spectrum synthesis}
\authorrunning{R. Hoppe et al.}

   \maketitle
\nolinenumbers

\section{Introduction}
Physically-realistic models of stellar spectra are central to many areas of astrophysics, including the determination of stellar chemical composition, interpretation of exoplanet transit light curves, and studies of stellar populations in the context of Galactic evolution \citep{GrevesseSauval1998, AsplundGrevesse2009, OhtaTaruya2005, EspinozaJordan2016, WatsonHansen2019, BergemannHoppe2026}.

Most studies in astronomy still rely on simple spectral models computed under the assumptions of 1D geometry, hydrostatic equilibrium (HE), and Local Thermodynamic Equilibrium (LTE). While the past decade has seen major advances in sophisticated radiative transfer methods \citep{StepanTrujilloBueno2013, HauschildtBaron2014, SteffenPrakapavicius2015, AmarsiNordlander2018}, their computational complexity continues to restrict their large-scale applicability. Modern astronomical surveys, such as 4MOST \citep{BensbyBergemann2019, BergemannBensby2026}, will observe millions of stars. These surveys require tens to hundreds of thousands of synthetic spectra to encompass the dimensionality of the observed data and the associated astrophysical diagnostics \citep{MeszarosAllendePrieto2012, SchonrichBergemann2014, EitnerBergemann2024, RozanskiTing2025, StormBergemann2025, StormBergemann2026}. For example, in the analysis of the APOGEE spectra, over 808 thousand atmospheric models and synthetic spectra were utilised in the calculations of stellar parameters and abundances \citep{MeszarosAllendePrieto2012}. Furthermore, high-resolution instruments like ESPRESSO and ANDES provide spectra of exquisite quality, enabling the tracing of stellar convection signatures through line bisectors \citep{Dravins2017a, Dravins2017b, Dravins2021}, which further necessitates the transition beyond 1D LTE frameworks.

This creates a discrepancy between the high-fidelity spectroscopic data provided by modern facilities and standard astrophysical codes, which are necessary to convert observed fluxes into fundamental stellar parameters, such as effective temperature, surface gravity, and detailed chemical abundances \citep{NissenGustafsson2018, JofreHeiter2019}. On the one hand, reliance on 1D LTE models severely limits the accuracy and precision of these derived parameters \citep{AsplundGrevesse2005, LindAmarsi2024, BergemannHoppe2026}. On the other hand, the use of such highly-parametrised 1D static stellar models precludes studying quantitatively and from first principles, many physical phenomena which are dynamic in nature. This is particularly evident in investigations of stellar convection \citep{Dravins2008, Maxted2018, DravinsLudwig2026} and its subsequent influence on the analysis of exoplanet spectra \citep{CanocchiLind2024, CanocchiMorello2024}.
To fully exploit the potential of upcoming astronomical observations, the community requires open-access 3D non-LTE (NLTE)-capable  stellar spectrum synthesis codes that rely on realistic, time-dependent stellar physics.

In this paper, we introduce \Code, an open-access code, designed to compute NLTE synthetic spectra using 3D radiation-hydrodynamics (RHD) stellar model atmospheres for AFGKM-type stars. It is an OpenMP-parallelised Fortran code integrated into the task-based \Dispatch framework \citep{NordlundRamsey2018, EitnerBergemann2024}. The code has been optimised and carefully tested with the goal to synthesise spatially-resolved and integrated flux spectra of stars using the \texttt{M3DIS} 3D RHD models, which share the same micro-physics. But our code is also compatible with other model atmospheres, such as 3D \Stagger \citep{MagicCollet2013, RodriguezDiazLagae2024}, \Muram \citep{Vogler2004, Vogler2005} and \Cobold \citep{Freytag2012}, as well as average $<$3D$>$ \citep{BergemannLind2012, KlevasKucinskas2016, BergemannCollet2017}, and classical 1D HE models \citep{Kurucz1970, GustafssonBell1975}. Most importantly, our code directly computes the statistical equilibrium for ions and molecules, and therefore self-consistent NLTE level populations for the full 3D cube can be employed in the multi-element 3D NLTE spectrum synthesis calculations for any stellar model in the AFGKM parameter space.

We in particular emphasise the high efficiency of the code. A broad-band detailed high-resolution spectrum (from 4000 to \SI{9000}{\angstrom}) of a typical G-type solar-metallicity star can be computed with a resolution of $R = 100\,000$ in $3.2$ seconds on a laptop (MacBook Pro, M1, 2020). This benchmark was done using a filtered line list with $325\,923$ atomic and molecular lines ($f_{\rm tol}=10^{-4}$). The line filtering method is outlined in Sect.~\ref{sec:line_opac}. Using the same line list, we computed the spectrum from a 3D snapshot with $30\times30\times96$ spatial points in $17$ minutes on a medium-sized compute node ($32\times$ Intel Xeon E5-2683, \SI{2.1}{GHz}). We further benchmarked the synthesis of departure coefficients for a small model atom such as oxygen (41 energy levels, 278 transitions, 3869 radiative frequency points) and on a large atom like Fe (607 levels, $13\,241$ transitions, $138\,604$ radiative frequency points) for the same star on both machines. The benchmarks are presented in Table~\ref{tab:benchmarks_comparison}.

We organise the paper as follows. In Sect.~\ref{sec:methods}, we detail the underlying microphysics, the numerical implementation and the radiative transfer methods for NLTE calculations and spectrum synthesis. In Sect.~\ref{sec:results}, we compare our results with established codes and discuss 3D NLTE effects on SEDs, Balmer lines and \spc{Mg}{1} abundances, alongside time-variable spectra. We present our conclusions in Sect.~\ref{sec:conclusions}.

\section{Methods} \label{sec:methods}
In this section, we outline the components and methods used in \Code in detail. Most critically, however, \Code supports NLTE treatment for bound-bound and bound-free transitions. The code computes the full statistical equilibrium for a system of energy states, which may represent a single ion or multiple ionisation stages of a chemical element or molecule. It employs the assumption of a trace ion, that is, no impact of the NLTE energy state populations on the model atmosphere structure or the equation of state (EOS). \Code comes with the NLTE model atoms and molecules from our previous dedicated studies, and these models are provided in Table~\ref{tab:nlte_atoms} and described in Sect.~\ref{sec:model_atoms}.

\Code relies on a wealth of atomic and molecular data, including ionisation/dissociation energies, partition functions, absorption cross-sections and damping constants. All of these parameters can be updated in dedicated text files and multiple input sets exist for backward compatibility. Input atmosphere models, NLTE atomic/molecular models, line lists, chemical abundances and isotopic fractions are flexible and can be customised to a given science problem.

In the following we will discuss our standard and carefully vetted, and therefore default inputs. Following the classification of \citet{PecautMamajek2013}, we tailor our treatment of the underlying microphysics to the atmospheric parameters of late-B and AFGKM-type stars. This includes all evolutionary phases covered by these spectral types: main-sequence, subgiants, red giants, AGB, and red supergiants. In particular, the line lists and opacity sources cover the ions and molecules that represent the conditions in those systems (Sect.~\ref{sec:EOS}).
According to our tests with \texttt{M3DIS} model atmospheres \citep{EitnerBergemann2024, EitnerBergemann2025} and 1D hydrostatic model atmospheres, such as \texttt{M1DIS}, \texttt{MAFAGS-OS} \citep{Grupp2004a, Grupp2004b} and \Marcs \citep{GustafssonEdvardsson2008}, our code is suitable for the calculation of synthetic stellar spectra in the following space of stellar parameters:
\begin{gather*}
\SI{2500}{K} \lesssim T_{\text{eff}} \lesssim \SI{8000}{K} \\
-1.0 \lesssim \log g \lesssim 5.5 \\
-10 \lesssim \text{[Fe/H]}\text{\footnotemark} \lesssim +1
\end{gather*}
\footnotetext{We have tested the code at metallicity $-10$, which effectively corresponds to metal-free atmospheres except for Li, and hence spectra only showing \spc{H}{1} and \spc{Li}{1} features.}Consequently, the code is intended for diagnostic spectroscopy applications within this space of stellar parameters.

\subsection{Equation of state} \label{sec:EOS}
The EOS was taken over from \MultiD, but features a novel implementation in modern Fortran 2018. Here, we present the code that runs and compiles with GNU Fortran Compiler (gfortran) 10 and Intel Fortran Compiler Classic (ifort) 2021.

We have added the option to use polynomial partition functions and extended the molecular equilibrium to handle an arbitrary number of molecules. The default files include data for the first 92 elements of the periodic table going up to \spc{C}{4}, \spc{N}{5}, \spc{O}{6}, \spc{Ne}{6}, \spc{Al}{4} and \spc{X}{3} for all other elements. The following $26$ molecules are included: non-polar molecules H$_2$, H$_2^+$, C$_2$, N$_2$, O$_2$, as well as dipolar molecules (those with a permanent dipole moment), H$_2$O, CH, NH, OH, CN, CO, NO, MgH, AlH, SiH, CaH, CrH, FeH, SiO, TiO, VO, ZrO, SiS, MgO, AlO and CaO.

The partition functions for all atoms are adopted from \citet{Irwin1981}. The partition functions and dissociation energies for all above-mentioned molecules, with the exception of H$_2$O, are from \citet{BarklemCollet2016}, which only provides data for diatomic molecules.
The H$_2$O partition function is adopted from \citet{BarberTennyson2006}, and it is consistent with the adopted H$_2$O line list and the dissociation energy is taken from \citet{Irwin1988}, as in \Turb, for backward compatibility. Molecular equilibrium constants are computed directly from these partition functions and dissociation energies, rather than interpolated in tabulated equilibrium constants. The partition functions entering the equilibrium constants are thereby identical to those used to convert number densities into level populations for the opacities, as required for consistency \citep{Irwin1981}. Neutral hydrogen level populations are determined using the occupation probability formalism \citep{HummerMihalas1988} following the implementation of \texttt{HLINOP} \citep{BarklemPiskunov2016}.

The reference set of solar abundances is taken from \cite{MaggBergemann2022}. In that study, the solar abundances were derived using different $<$3D$>$ solar model atmospheres and NLTE line formation for C, N, O, Mg, Si, Ca, Fe and Ni computed using \Turb and \texttt{MULTI2.3} \citep{GerberMagg2023}, with model atoms of the kind described in Sect.~\ref{sec:model_atoms}, and this dataset also includes NLTE values of abundances of other chemical elements (Na, Al, S, K, Sc, Ti, V, Cr, Mn, and Co) adopted from previous studies. We use this solar data set because it maximises internal consistency with our model atmosphere calculations with the \texttt{M3DIS} code \citep{EitnerBergemann2024, EitnerBergemann2025}, which are the standard inputs to our code, sharing the same micro-physics as \Code for full consistency. We note that these solar abundances fulfil helioseismic constraints on the solar interior \citep{SerenelliBasu2009, MaggBergemann2022} and latest comprehensive neutrino results for the pp chain and CNO cycle \citep{AppelBagdasarian2022, BasilicoBellini2023, Gonzalez-GarciaMaltoni2024}, but the user may flexibly choose any other solar composition because it is tabulated in a single text file. The isotopic weights and fractions are compiled from \cite{BergemannLodders2026}.

\subsection{Opacity package}
The continuum opacities employed by \Code are provided alongside their references in Table~\ref{tab:cont_opac}.

We have updated all absorption cross-sections to match those used by \Marcs and \Turb. The input formats and interpolation procedures are based on the Uppsala Opacity Package \citep{GustafssonBell1975} for backward compatibility. Additionally, we have implemented the occupation probability formalism for \spc{H}{1} bound-free transitions following \texttt{HLINOP} \citep{BarklemPiskunov2016}.

\begin{table}
\newcommand{\rom}[1]{\MakeUppercase{\romannumeral #1}}
\centering
\caption{Default continuum opacities and their sources.}
\label{tab:cont_opac}
\begin{tabularx}{\linewidth}{l l X}
\toprule
Species   &  Type &  Source \\
\midrule
H                               &  bf, ff   & \cite{Karzas1961} \tablefootmark{a} \\
H$^{-}$                         &  bf       & \cite{McLaughlin2017} \\
He, C, N, O, Mg                 &  bf       & TOPbase \citep{Seaton1994} \\
 Al, Si, Ca, Fe                 &  bf       & TOPbase \citep{Seaton1994} \\
CH, OH                          &  bf       & \cite{Kurucz1987} \\
H$^{-}$                         &  ff       & \cite{Bell1987} \\
He, C, Mg, Si                   &  ff       & \cite{Peach1970} \\
other metals                    &  ff       & \cite{Peach1970} \\
He$^{-}$                        &  ff       & \cite{John1994} \\
H$_{2}^{-}$, O$^{-}$, CO$^{-}$  &  ff       & \cite{John1975b, John1975a} \\
H$_{2}$O$^{-}$                  &  ff       & \cite{John1975b, John1975a} \\
N$^{-}$                         &  ff       & \cite{RamsbottomBell1992} \\
H$_{2}^{+}$                     &  ff       & \cite{Mihalas1965} \\
H + H                           &  cia      & \cite{Doyle1968} \\
H + He                          &  cia      & \cite{GustafssonFrommhold2001} \\
H$_{2}$ + H                     &  cia      & \cite{GustafssonFrommhold2003} \\
H$_{2}$ + H$_{2}$               &  cia      & \cite{BorysowJorgensen2001} \\
H$_{2}$ + He                    &  cia      & \cite{JorgensenHammer2000} \\
e$^-$                           &  scat     & \cite{Mihalas1978} \\
H                               &  scat     & Dalgarno as in \cite{Gingerich1964} \\
H$_{2}$                         &  scat     & \cite{DalgarnoWilliams1962} \\
He                              &  scat     & \cite{BuesWehrse1976} \\
\bottomrule
\end{tabularx}
\tablefoot{Opacity types are bound-free (bf), free-free (ff), collision-induced or quasi-molecular absorption (cia) and scattering (scat). These sources have been compiled by the authors of \texttt{MARCS} and \texttt{TURBOSPECTRUM} \citep{GustafssonEdvardsson2008, Plez2012, GerberMagg2023}.
\tablefoottext{a}{Modified using the occupation probability formalism following \cite{BarklemPiskunov2016}.}}
\end{table}
 
\subsection{Line opacities} \label{sec:line_opac}
The opacities due to radiative bound-bound transitions are computed using atomic and molecular line lists sorted by isotope/isotopologue. The code reads the \Turb format natively as well as a second simplified format presented in Table~\ref{tab:llist_demo}. Common line list formats, including \texttt{VALD}, \texttt{MOOG} and \texttt{LINEMAKE} can be converted into the required format using Python scripts which are shipped as part of the distribution. However, we strongly recommend our carefully curated line lists published alongside our work in \citet{GerberMagg2023} for scientific use\footnote{The recommended line lists are available under \url{https://keeper.mpdl.mpg.de/d/6eaecbf95b88448f98a4/}}.

The linear absorption coefficient $\chi_\nu$ and the emissivity $\eta_\nu$ are computed according to equations~\ref{eq:chi} and \ref{eq:eta}.

\begin{align}
    \label{eq:chi}
    \chi_{\nu} &= \frac{h\nu}{4\pi} \phi_{\nu} n_i^{\text{LTE}} B_{ij}\,(b_i - \beta_\nu\,b_j ), \\
    \label{eq:eta}
    \eta_\nu &= \frac{2h\nu^3}{c^2} \frac{h\nu}{4\pi} \phi_{\nu} n_i^{\text{LTE}} B_{ij}\,\beta_\nu\,b_j,
\end{align}

where $\phi_\nu$ is the line profile, $n_i^{\rm LTE}$ is the number density of the lower energy level, $B_{ij}$ is the absorption Einstein coefficient, $b_i = n_i / n_i^{\rm LTE}$ is the departure coefficient and $\beta_\nu = \exp(-h\nu/kT)$ is the Boltzmann factor. The other symbols have their usual meanings. The assignment of departure coefficients to individual lines is described in Sect.~\ref{sec:model_atoms}. For more details on the chosen formalism see Appendix~\ref{sec:line_kernel}.

We avoid computationally expensive calculations in the distant line wings, where contributions to the total opacity are negligible. The code initiates the calculation of the line opacity $\chi_{\rm line}$ at $\lambda_0$ and verifies if $\chi_{\rm line} > f_{\rm tol} \, \chi_{\rm cont}$, where $f_{\rm tol}$ is a user-defined tolerance and $\chi_{\rm cont}$ is the continuum opacity at $\lambda_0$. Each wing is then computed by stepping outwards, terminating the process once the line opacity falls below the threshold $f_{\rm tol} \, \chi_{\rm cont}$. Where the line opacity already accumulated at $\lambda_0$ exceeds $\chi_{\rm cont}/f_{\rm tol}$, the threshold is raised to $f_{\rm tol}^2$ times that opacity, so that weak lines inside saturated band cores are skipped. All calculations in this paper use the default $f_{\rm tol}=10^{-4}$. The code records $\max(\chi_{\rm line} / \chi_{\rm cont})$ of each line over all grid points and writes a filtered line list after each synthesis, which contains only the lines for which this ratio exceeds $f_{\rm tol}$. The filter therefore depends on the physical conditions of the model atmosphere and on the wavelength range. We use it to reduce the master line lists using a 1D model atmosphere before running a 3D synthesis. The default $f_{\rm tol}=10^{-4}$ is the value adopted in \Turb \citep{Plez2012, GerberMagg2023} and \texttt{SYNSPEC} \citep{HubenyAllendePrieto2021}, and a factor of three below the default of \texttt{Korg} \citep{WheelerAbruzzo2023}. At this tolerance, the temperature differences between 1D and 3D models are not critical for the line selection in spectroscopic diagnostics. Diagnostic lines lie well above the threshold, and lines close to it only contribute to a weak pseudo-continuum that is removed by local renormalisation of both observed and synthetic spectra.

\begin{table}
\footnotesize
\centering
\caption{Example of a line list entry.}
\label{tab:llist_demo}
\setlength{\tabcolsep}{4pt}
\begin{tabular}{cccccccc}
\toprule
$\lambda_0$ [\AA] & E$_i$ [eV] & $\log(gf)$ & $\gamma_{\rm vdw}$ & $\gamma_{\rm rad}$ & $\gamma_{\rm stk}$ & i & j                      \\
\midrule
7771.94 &  9.146 & 0.369 & 453.234 & 3.89e+07 & 0 & 0 & 0  \\
\bottomrule
\end{tabular}
\tablefoot{The columns are the central wavelength $\lambda_0$, lower excitation potential E$_i$, log(gf) value and damping parameters $\gamma_{\rm vdw}$, $\gamma_{\rm rad}$, and $\gamma_{\rm stk}$ for van der Waals, radiative, and Stark broadening, respectively. Additionally, indices i and j may be specified to connect the lower and upper energy levels to precomputed departure coefficient data.}
\end{table}
 
Table~\ref{tab:llist_demo} provides an example of an entry in a line list file for a radiative bound-bound transition. The minimal requirement is the central wavelength $\lambda_0$, lower excitation potential E$_i$, the transition probability $\log(gf)$, a product of an f-value and the statistical weight of the lower energy level. All other columns are optional, so that $\gamma_{\rm vdw}$ is only read if a 4th column is provided and $\gamma_{\rm rad}$ if there is a 5th column and so on.

Intrinsic line broadening is included through a Voigt profile resulting from the convolution of the Gaussian thermal (and optionally microturbulence) broadening profile with the radiative and collisional broadening Lorentz profile. The damping coefficients are computed according to various implemented recipes, which typically scale with $N_{\rm \spc{H}{1}}$ in the case of van der Waals broadening and $N_e$ in the case of Stark broadening (see Appendix~\ref{sec:damping}).

For \spc{H}{1} lines we use dedicated profiles that include linear Stark broadening. For the Balmer lines up to $n=20$ and the Paschen lines up to $n=21$ we interpolate the tabulated profiles of \citet{StehleHutcheon1999}, computed with the Model Microfield Method, which includes ion dynamics. For the Brackett lines we interpolate the VCS \citep{VidalCooper1973} profiles of \citet{Lemke1997}, whose tables were taken over from the SIU code \citep{Reetz1991}. The tables include thermal Doppler broadening. The other broadening mechanisms are added to the tabulated profile as a single Lorentzian outside the line core. These are radiative damping, broadening by neutral helium, and self-broadening by neutral hydrogen, the latter from \citet{BarklemPiskunov2016} for H$\alpha$, H$\beta$ and H$\gamma$ and from \citet{AliGriem1966} for all other lines. For lines outside the tables, including the Lyman series, we represent quasistatic ion perturbations using a Holtsmark profile, and we model electron impact effects as an additional Lorentzian contribution. Furthermore, we transition from a convolved Voigt profile in the line core to a direct summation of the independent Gaussian, Lorentzian, and Holtsmark profiles in the line wings following the implementation by \cite{BarklemPiskunov2016}. An application to the Balmer series of a metal-poor red giant is shown in Sect.~\ref{sec:balmer}.

\subsection{Input model atmospheres}
\Code can take different types of stellar model atmospheres as input, including 1D, 2D, and 3D models.

The basic information required from the 3D cube are the spatial $x,y,z$ coordinates, velocities $v_x,v_y,v_z$, temperature $T$ and one of the following: electron number density $n_e$, electron pressure $p_e$ or mass density $\rho$. The missing quantities out of $n_e, p_e$ and $\rho$ are determined via the EOS. The code works in CGS units. Natively supported formats for 3D radiation-hydrodynamics model atmosphere consist of \texttt{M3DIS} \citep{EitnerBergemann2024}, \texttt{STAGGER} \citep{MagicCollet2013}, \texttt{STAGGER2.0} \citep{RodriguezDiazLagae2024} and \texttt{MULTI3D} \citep{LeenaartsCarlsson2009}.

The code also takes standard 1D hydrostatic model atmospheres as input. In contrast to 3D models, the 1D models may also be tabulated on an optical depth scale $\log_{10}(\tau_{\rm 500\,nm})$ and may include a depth-dependent microturbulence $v_{\rm mic}$ in \si{\km\per\s}. Optical depth scales are converted to height scales via the EOS as the code only uses height internally. The supported 1D formats are: \texttt{MARCS} \citep{GustafssonEdvardsson2008}, \texttt{MAFAGS-OS} binary files \citep{Grupp2004a, Grupp2004b} and a custom text format ($z\,\text{or}\,\log_{10}(\tau_{\rm 500\,nm}),T,p_e,\rho,v_{\rm mic}$).

\subsection{Radiative transfer}
The code employs a short-characteristics solver \citep{MihalasAuer1978, Kunasz1988} during NLTE iterations and a long-characteristics solver \citep{JonesSkumanich1973, Jones1973} for spectrum synthesis. Both linear and monotonic quadratic solvers are provided adopting a plane-parallel configuration for all calculations.

In the current version, the boundary condition at the top of the atmosphere is zero incoming radiation ($I=0$). At the bottom boundary, we extrapolate the Planck function assuming $I_\nu=B_\nu$. Horizontal boundary conditions are periodic, following the properties of 3D radiation-hydrodynamics model atmospheres. Consequently, the atmosphere is not split horizontally to avoid circular dependencies between RT patches.
\begin{table}
  \centering
  \caption{Adopted sources of NLTE model atoms and molecules distributed with the code.}
  \label{tab:nlte_atoms}
  \begin{tabular}{ll}
    \toprule
    Atom & Literature study \\
    \midrule
    H  & \citet{MashonkinaZhao2008} \\
    O  & \citet{BergemannHoppe2021} \\
    Na & \citet{EzzeddineMerle2018} \\
    Mg & \citet{BergemannCollet2017} \\
    Al & \citet{EzzeddineMerle2018} \\
    Si & \citet{BergemannKudritzki2013}\tablefootmark{a} \\
    Ca & \citet{MashonkinaKorn2007}\tablefootmark{b} \\
    Ti & \citet{Bergemann2011} \\
    Cr & \citet{BergemannCescutti2010} \\
    Mn & \citet{BergemannGallagher2019} \\
    Fe & \citet{BergemannLind2012}\tablefootmark{b} \\
    Co & \citet{BergemannPickering2010}\tablefootmark{c} \\
    Ni & \citet{BergemannHoppe2021}\tablefootmark{d} \\
    Sr & \citet{BergemannHansen2012} \\
    Ba & \citet{GallagherBergemann2020} \\
    Y  & \citet{StormBergemann2023} \\
    Eu & \citet{StormBarklem2024} \\
    CH & \citet{HoppeBergemann2026} \\
    \bottomrule
  \end{tabular}
  \tablefoot{
    \tablefoottext{a}{With Si+H collisional data from \citet{BelyaevYakovleva2014} and updates to the radiative data as described in \citet{MaggBergemann2022}.}
    \tablefoottext{b}{Updates described in \citet{SemenovaBergemann2020}.}
    \tablefoottext{c}{With Co+H collisional data from \citet{YakovlevaBelyaev2020}.}
    \tablefoottext{d}{With Ni+H collisional data from \citet{VoronovYakovleva2022}.}
  }
\end{table}
 
\subsection{NLTE model atoms} \label{sec:model_atoms}
The NLTE model atoms can be used in the same format as those developed for \Multi \citep{Carlsson1986} and \MultiD \citep{LeenaartsCarlsson2009}. Our standard version of the code already includes the following published well-tested NLTE model atoms for 17 elements: H, O, Na, Mg, Al, Si, Ca, Ti, Cr, Mn, Fe, Co, Ni, Sr, Ba, Y, Eu. See Table~\ref{tab:nlte_atoms} for references. These NLTE model atoms were released as part of \Turb and new models, including CH, can be retrieved from the same open-access database maintained by MPIA\footnote{\url{https://keeper.mpdl.mpg.de/d/6eaecbf95b88448f98a4/}}.

Where a level of the model atom combines several fine-structure levels or terms, all lines of these levels share its departure coefficient. This assumes that the combined levels are populated relative to one another according to the Boltzmann distribution. However, we note that this is a very rare case, as for all atomic models (Table~\ref{tab:nlte_atoms}) used in the code and released by our group, direct fine-structure levels are used in the model atom for the energies that are representative of UV, optical, and IR transitions. Only states close to the ionisation limit are combined into terms or superlevels, which in the Fe and Ni model atoms applies above about 6.8 \citep{BergemannLind2012, MaggBergemann2022} and \SI{6.2}{\electronvolt} \citep{BergemannHoppe2021}, respectively. The only exception to our standard approach is the CH molecule \citep{HoppeBergemann2026}, whose 1981 states were grouped by excitation energy into 294 superlevels using K-means clustering.

The departure coefficients computed by the statistical equilibrium solver (Sect.~\ref{sec:se_solver}), either in 1D or in 3D, are stored in a separate file and can be associated with individual spectral lines in a line list. This is done via the use of energy state indices $i$ and $j$, which can be supplied in the line list file. These indices refer to the levels of a specific model atom, so a line list has to be matched to the model atom it is used with.

For line lists without these indices, the code can optionally assign them with the procedure developed for \Turb and \texttt{TSFitPy} \citep{GerberMagg2023}, which is not ideal, but provides a starting point for further improvements. See a more careful and rigorous approach used in the SIU code \citep[e.g.][]{KornShi2003, BergemannPickering2010, MashonkinaZhao2008}. In the \texttt{TSFitPy} approach, a line takes the levels of the model atom transition within \SI{0.02}{\angstrom} of its wavelength whose lower-level energy agrees to within a factor of 1.0005, and otherwise the nearest level within that tolerance. Lines without a match remain in LTE. Indices given in the line list are always kept. We emphasise that no attempt is made in this paper to provide a more physically correct and elaborate approach, in which the upper and lower energy levels, including their properties (such as total angular momentum, energies) are used to ensure the correct states are identified with the radiative bound-bound transition, and this is essential for any careful stellar spectroscopic NLTE analysis.

\subsection{Statistical equilibrium solver} \label{sec:se_solver}
We evaluate convergence by tracking the maximum relative change in the departure coefficients between iterations, $e_i = b_i / b_i^{\rm old} - 1$. By default, we consider the solution converged when the maximum absolute value of $e_i$ falls below $10^{-3}$.

One advantage of the task-based parallelisation scheme of \Dispatch is the possibility to apply asynchronous updates to individual subregions. We give the user the option to exclude subregions from the iterative updates once their populations converge below an independent, customisable threshold. Optically thick subregions typically converge within a few iterations and are unaffected by changing populations in upper atmospheric layers. Excluding them reduces the computational cost of subsequent lambda iterations, as the solver focuses strictly on the optically thin regions that require further iterations to reach statistical equilibrium.

\MultiD originally used the full-preconditioning scheme \citep[][Sect.~2.3]{RybickiHummer1992} of the rate equations, which eliminates cross-terms in the transition matrix when overlapping lines are present. We note that preconditioning serves strictly to accelerate the convergence rate of the iterative solver and does not alter the final converged solution. Because calculating all cross-terms is computationally expensive and generally not required to achieve convergence, we implement the simpler auto-preconditioning scheme (Sect.~2.4 of the same paper). This method isolates the feedback in the radiative rates of each line with itself, which reduces memory and computation costs while yielding identical physical results to the full scheme. Other 3D NLTE codes, such as \Balder \citep{AmarsiBarklem2019}, also use this approach. Optionally, Ng acceleration \citep{Ng1974} can be enabled, which may significantly speed up the convergence.

\subsection{Atmosphere projection} \label{sec:projection}

To reduce numerical interpolation errors associated with evaluating inclined rays in downsampled 3D atmospheres, we implement a pre-tilting atmosphere projection scheme for our long characteristics solver. This method has previously been suggested by \cite{PeckCriscuoli2017} for interpolating stellar atmosphere models. They have discussed and tested the concept of angle-aligning the grid prior to computing radiative transfer to reduce intensity diffusion in short-characteristics solvers. While long-characteristics solvers do not suffer from diffusion due to interpolation in the intensity, they do suffer from diffusion when interpolating the source function and especially the opacity onto the inclined photon paths. Interpolating the hydrodynamic quantities prior to downsampling onto tilted grids may also introduce interpolation errors, but it is still favourable, since one may interpolate in the full-resolution snapshot, rather than in the already downsampled version used during the radiative transfer calculations. The same method is in fact briefly mentioned in \citet{HayekAsplund2011} as part of the 3D LTE radiative transfer code \texttt{SCATE}; however, no comparisons against other methods are presented in said paper. This pre-tilting approach only comes with one additional complexity in the case of NLTE line formation. Since we typically compute the departure coefficients on a downsampled non-tilted grid, we then have to interpolate the departure coefficients onto the ray-aligned grid.

\begin{table}
\centering
\footnotesize
\setlength{\tabcolsep}{5pt}
\caption{Stellar parameters of the employed model atmospheres.}
\label{tab:stellar_parameters}
\begin{tabularx}{\linewidth}{X c c c c c}
\toprule
Star & $T_{\rm eff}$ & $\log g$ & [Fe/H] & $v_{\rm mic}$\tablefootmark{a} & 3D Model \\
\midrule
Sun  & 5777 & 4.4 & 0.0 & 1  & \Stagger \\
M Dwarf & 4000 & 4.5 &  0.0 & 1  & \Mdis \\
Metal-poor RGB & 5000 & 2.0 & -3.0 & 2  & \Mdis \\
Metal-poor RGB\tablefootmark{b} & 5000 & 2.0 & -3.0 & 2  & \Stagger \\
Metal-poor Turn-off  & 6500 & 4.0 & -2.0 & 1  & \Stagger \\
Metal-poor Turn-off\tablefootmark{c}  & 6500 & 4.0 & -3.0 & 1  & -- \\
Red Supergiant & 4000 & 1.0 &  0.0 & 5  & \Cobold \\
\bottomrule
\end{tabularx}
\tablefoot{
\tablefoottext{a}{$v_{\rm mic}$ was only used for the 1D spectrum synthesis utilising corresponding \Marcs model atmospheres.}
\tablefoottext{b}{In the validation against \MultiD we employed the \Stagger model as the \Mdis format is not supported in the legacy code.}
\tablefoottext{c}{Only used with the 1D \Marcs model in the validation against \Multi.}
}
\end{table}
 
\begin{table}[ht]
\centering
\caption{Performance benchmarks for various synthesis tasks.}
\label{tab:benchmarks_comparison}
\begin{tabularx}{\linewidth}{l l @{\hspace{5pt}} r l @{\hspace{5pt}} r}
\toprule
& \multicolumn{2}{X}{Laptop (M1)} & \multicolumn{2}{X}{Node (32 CPUs)} \\
\cmidrule(lr){2-3} \cmidrule(l){4-5}
Task & Time & \hspace{-1em}Memory & Time & \hspace{-1em}Memory \\
\midrule
Hi-res optical 1D   & 3.2 s   & 0.5 GB & 5.0 s   & 0.8 GB \\
Hi-res optical 3D   & ---     & ---    & 17 min  & 28 GB  \\
Dep. coeff. 1D (O)  & 0.3 s   & 0.1 GB & 1.3 s   & 0.4 GB \\
Dep. coeff. 1D (Fe) & 19 s    & 1.3 GB & 18 s    & 1.8 GB \\
Dep. coeff. 3D (O)  & 12 min & 1.5 GB & 6.2 min & 2.7 GB \\
Dep. coeff. 3D (Fe) & ---     & ---    & 10 h   & 39 GB  \\
\bottomrule
\end{tabularx}
\tablefoot{Times are wall-clock times, and memory is the peak resident set size measured by the operating system.}
\end{table}
 
\section{Results and discussion} \label{sec:results}

\begin{figure*}[!ht]
    \centering
    \includegraphics[scale=\pltscale]{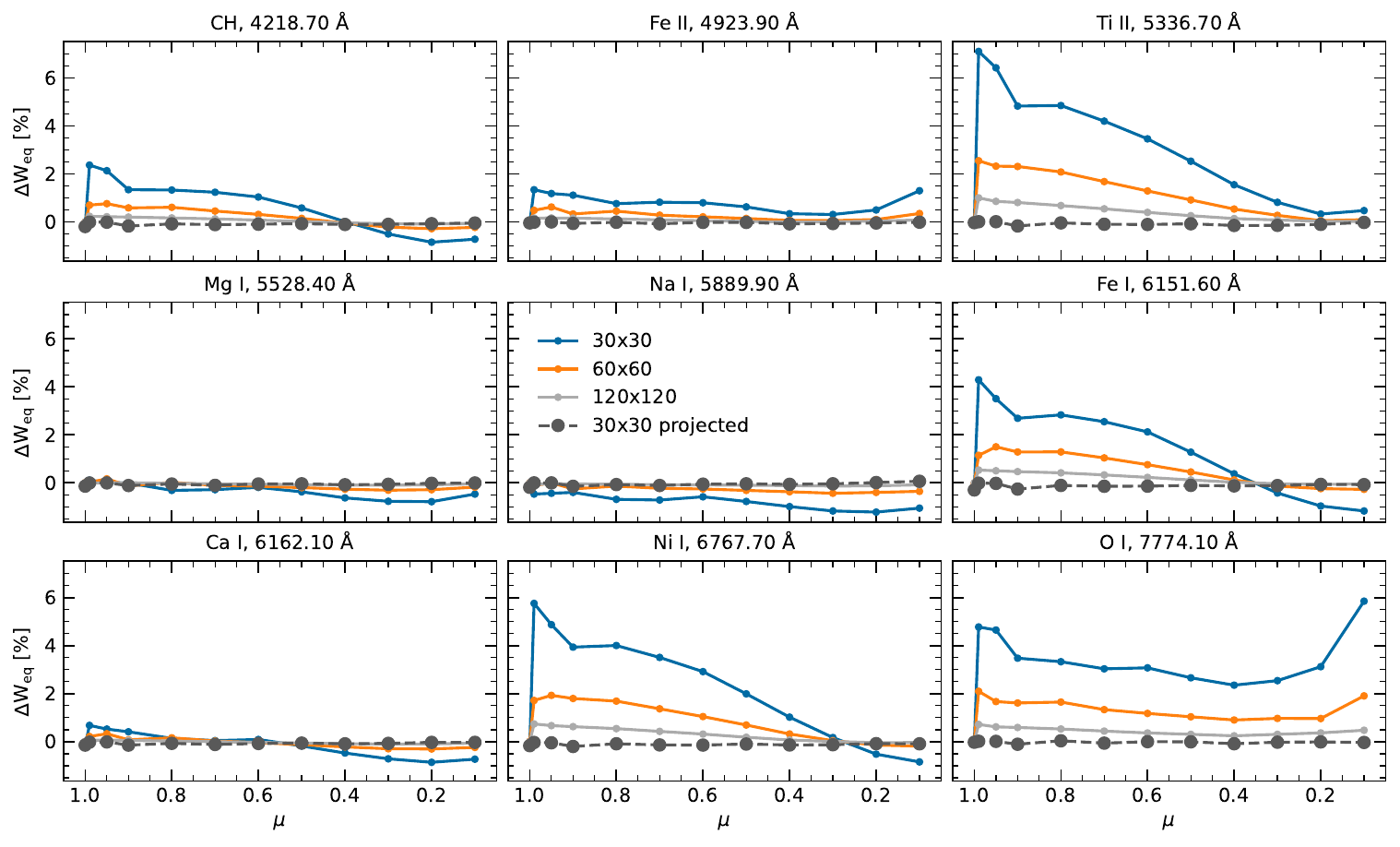}
    \caption{Equivalent widths as a function of inclination angle $\mu$ for varying horizontal resolutions, comparing results with and without pre-tilting of the hydrodynamical model atmosphere snapshots. The values are expressed relative to the equivalent widths of the full-resolution ($240 \times 240$) result. In all cases, the pre-tilted (projected) approach reproduces the full-resolution equivalent width to within $0.5\%$, despite its relatively low horizontal resolution of $30 \times 30$.}
    \label{fig:projection_weqs}
\end{figure*}

\subsection{Atmosphere projection validation}

In order to test the properties of the atmosphere projection method, we have selected 9 spectral lines of various species, ionisation stages, lower excitation potentials and oscillator strengths, hence a selection of lines covering a broad range of formation depths within the photosphere. The employed line data are listed in Table~\ref{tab:benchmark_lines}. We compute 3D LTE line profiles for 12 different inclinations employing one solar atmosphere snapshot. We carry out this calculation using the full-resolution snapshot with horizontal resolution of 240$\times$240 points and progressively downsampled versions of the same snapshot. During the radiative transfer, the source function and opacity are linearly interpolated onto the photon path in this setup. Alternatively, we compute the same line profiles, but by interpolating the hydrodynamic quantities prior to downsampling and prior to the radiative transfer calculation onto tilted grids with a horizontal resolution of 30$\times$30 points.

We present the results in Fig.~\ref{fig:projection_weqs}, where we show the error in the equivalent widths $W_{\rm eq}$ of each spectral line with respect to the full-resolution solution. We show
\begin{equation}
    \Delta W_{\rm eq} = \frac{W_{\rm eq, n_x\times n_y}}{W_{\rm eq, 240\times240}} - 1
\end{equation}
as a function of inclination angle $\mu=\cos\theta$, where $\theta$ is the inclination of the ray with respect to the z-axis. $\mu$=1 corresponds to the vertical ray, where no interpolation is needed  as the ray aligns with the grid. Any differences in $W_{\rm eq}$($\mu$=1) are from probing the stellar granulation pattern more or less often. Interestingly, even a downsampled snapshot with a resolution of 30$\times$30 shows very little deviation in $W_{\rm eq}$($\mu$=1) compared to the full-resolution result. However, for slightly inclined rays, i.e. $\mu$=0.99 and $\mu$=0.95, $W_{\rm eq}$ can deviate up to $\sim7\%$, in the case of the \spc{Ti}{2} line. We also note that the $W_{\rm eq}$ tends to be overestimated near the vertical, but in some cases can also be underestimated, as is seen in the \spc{Na}{1} line.

As expected, the interpolation error is reduced at higher resolution, with typically $\Delta W_{\rm eq} < 1\%$ at a resolution of 120$\times$120. In contrast, the full-resolution $W_{\rm eq}$ is reproduced to within 0.5\% for every benchmark line when using a reduced horizontal resolution of 30$\times$30 in combination with the projection approach. This clearly demonstrates the superior efficiency of this method as the horizontal resolution has a large impact on the computational cost of the radiative transfer calculation. For the interpolation of $b_i$ in NLTE, we found that linear interpolation in $\log b_i$ with subsequent normalisation of the populations to reproduce the total number density of the analysed species reduces interpolation errors compared to linear interpolation in $b_i$.

\begin{figure}[!ht]
    \centering
    \includegraphics[width=\linewidth]{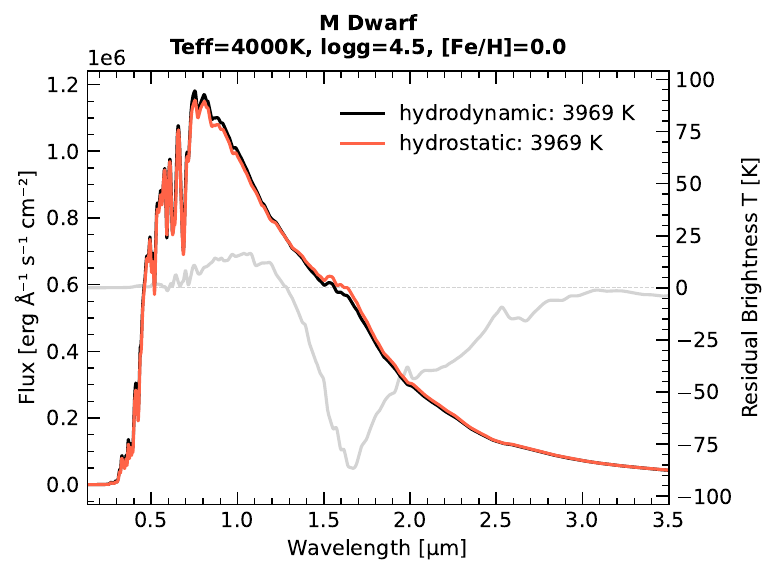}
    \includegraphics[width=\linewidth]{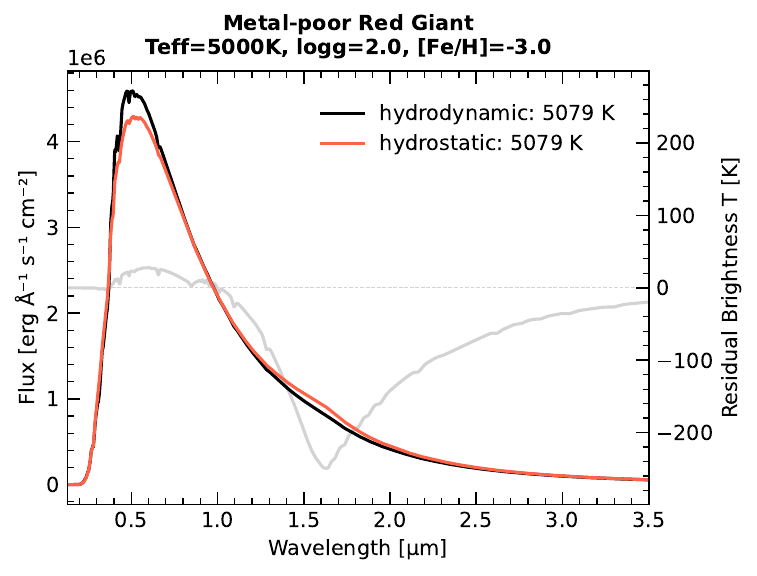}
    \includegraphics[width=\linewidth]{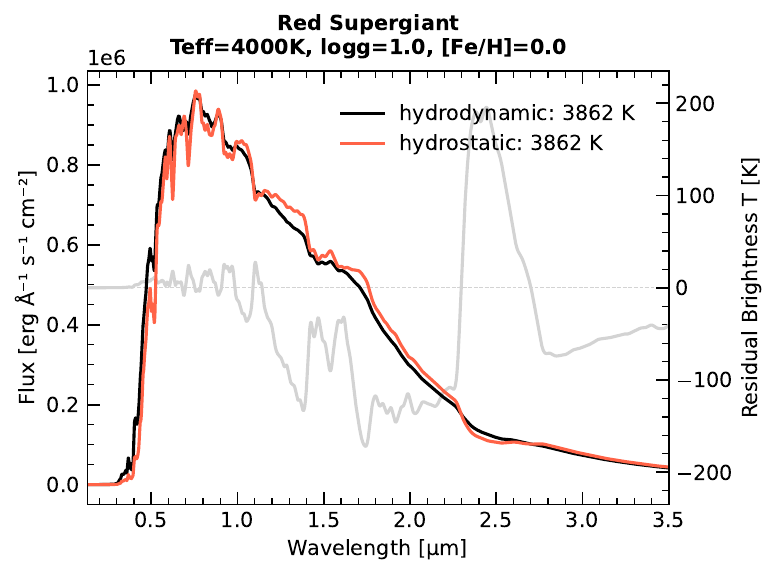}
    \caption{SED test calculation of a typical M dwarf, metal-poor red giant and red supergiant computed in 1D and 3D LTE using \Code. The spectra have been convolved to a resolution of $R = 100$ for easier visual comparison. The legend states the effective temperature of the spectra from direct integration.}
    \label{fig:sed_3d}
\end{figure}

\begin{figure}[!ht]
    \centering
    \includegraphics[scale=\pltscale]{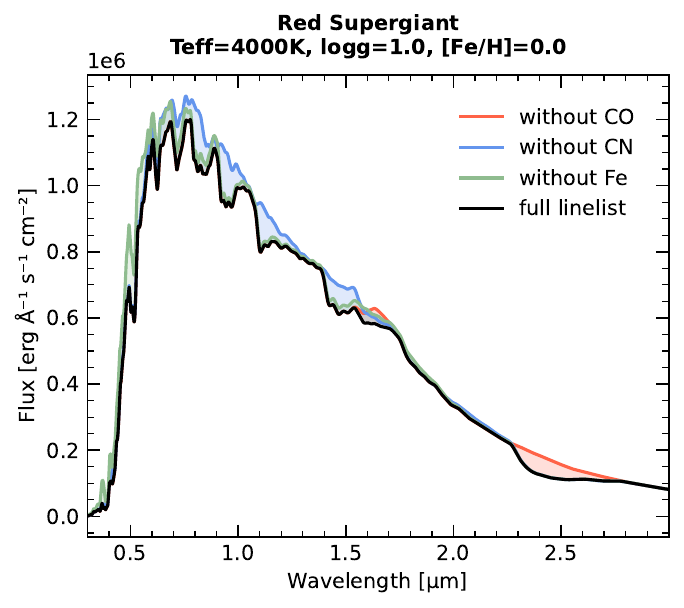}
    \caption{SED test calculation of a typical red supergiant with and without CO, CN and Fe lines. This comparison was done using a hydrostatic \Marcs model atmosphere.}
    \label{fig:sed_species}
\end{figure}

\subsection{Code validation}
We aim to employ \Code both as a reliable source of NLTE departure coefficients and as a spectrum synthesis code for 1D and 3D stellar atmospheres. This requires validation against other established codes commonly used for either application.

The successful validation of our code against \Multi, \MultiD and \Turb is presented in Appendix~\ref{sec:validation}. First we compare 1D LTE and 1D NLTE line profiles against \Multi to validate the formal radiative transfer solver and ensure that the statistical equilibrium solver works reliably for simple atoms, Mg from \citet{BergemannCollet2017}, as well as complex model atoms, Fe from \citet{BergemannLind2012}. The utilised atomic line data for the validation is presented in Table~\ref{tab:nlte_lines}. Then we compare 3D LTE and 3D NLTE line profiles against \MultiD to test the atmospheric sampling, the short-characteristics and the long-characteristics solver in the case of 3D radiative transfer.
And finally, we compare 1D LTE spectral energy distributions (SEDs) of multiple stellar types against \Turb to demonstrate the applicability of the code to various stellar regimes. Table~\ref{tab:line_lists} specifies the line list sources used for all SEDs presented in this paper. The molecular line lists were taken from the \Turb distribution\footnote{\url{https://github.com/bertrandplez/Turbospectrum_NLTE/tree/master}}, because \Turb is the validation target for our 1D SED calculations (Appendix~\ref{sec:turbo_validation}). We have not adopted line lists from the ExoMol database \citep{TennysonYurchenko2024}, because ExoMol does not provide broadening by atomic hydrogen and a validation of its data for stellar atmospheres is beyond the scope of this paper. For specific applications, users can convert ExoMol or other custom line lists to one of the supported formats.

\begin{table}
\caption{Sources of atomic and molecular line lists used for the SED calculations presented in this paper.}
\label{tab:line_lists}
\centering
\begin{tabular}{l l}
\toprule
Species & Reference \\
\midrule
\multirow{2}{4em}{Atoms} & \texttt{VALD} \citep[][and refs. therein]{PiskunovKupka1995} \\
    & \texttt{VALD-2} \citep[][and refs. therein]{KupkaRyabchikova2000} \\
H$_2$O & \citet{BarberTennyson2006} \\
CH  & \citet{MasseronPlez2014} \\
NH  & \citet{Kurucz2011} \\
\multirow{2}{4em}{OH}  & \citet{GoldmanSchoenfeld1998}, \\
    & \texttt{HITRAN} \citep{GordonRothman2017} \\
MgH & \citet{SkoryWeck2003} \\
AlH & \citet{Plez1998, Plez2012} \\
SiH & \citet{Kurucz2011} \\
CaH & \citet{Plez1998, Plez2012} \\
CrH & \citet{BurrowsRam2002} \\
FeH & \citet{Plez2012} \\
C$_2$ & \citet{QuerciQuerci1971} \\
CN  & \citet{HedrosaAbia2013, Plez2012} \\
CO  & \citet{Goorvitch1994, LiGordon2015} \\
SiO & \citet{LanghoffBauschlicher1993} \\
TiO & \citet{Plez1998, Plez2012} \\
VO  & \citet{Plez1998, Plez2012} \\
ZrO & \citet{Plez1998, Plez2012} \\
SiS & \citet{CamiSloan2009} \\
\bottomrule
\end{tabular}
\end{table}
 
In the following sections we present the 3D LTE SEDs and their differences from the 1D counterparts (Sect.~\ref{sec:sed_3d}). We then show a synthetic near-UV spectrum computed with NLTE departure coefficients for \spc{H}{1} and \spc{Fe}{1}, and examine how 3D granulation affects the wings of the Balmer series (Sect.~\ref{sec:balmer}). We also investigate the time variability of line profiles across the 3D snapshot series, which 1D hydrostatic models cannot produce (Sect.~\ref{sec:time_series}). Finally, we compute 3D NLTE$-$1D LTE and 1D NLTE$-$1D LTE abundance corrections for three \spc{Mg}{1} lines in a metal-poor turn-off atmosphere and compare them with literature values (Sect.~\ref{sec:mg_corrections}). Table~\ref{tab:stellar_parameters} lists the properties of the model atmospheres.

\subsection{3D SEDs} \label{sec:sed_3d}
For the hydrodynamic vs. hydrostatic (3D $-$ 1D) comparison of the spectral energy distributions (SEDs), we interpolated spectra from multiple 1D \Marcs model atmospheres spanning a range of effective temperatures in order to match the \Teff of the 3D snapshots. For all models, \Teff was computed by direct integration of the emergent flux over $\sim\SI{100000}{}$ wavelengths in the range \SI{1300}{}--\SI{200000}{\angstrom}, using exactly the same wavelength grid as adopted in the \Marcs code (flx\_wavelengths.vac)\footnote{\url{https://marcs.oreme.org}}. The spectra are shown convolved to a resolving power of $R = 100$.

Despite differences in stellar parameters, the three cases presented in Fig.~\ref{fig:sed_3d} display a consistent trend. Compared to 1D counterparts, 3D SEDs are systematically cooler at the H$^-$ opacity minimum near \SI{1.6}{\micro\metre} and moderately hotter in the optical range (\SIrange{0.4}{0.8}{\micro\metre}). Noticeable differences are seen in the red supergiant, in particular for the red CN band system (\SIrange{0.8}{2.0}{\micro\m}), the first overtone band of CO (\SIrange{2.3}{2.7}{\micro\m}) and optical Fe lines (\SIrange{0.3}{0.8}{\micro\m}). For demonstration purposes we show the impact of these species on the SED of a red supergiant using a 1D \Marcs models atmosphere in Fig.~\ref{fig:sed_species}. The CO band, for example, is considerably weakened in the 3D model, counteracting the offset caused by the H$^-$ opacity minimum. The 3D $-$ 1D discrepancies are generally smallest in the M dwarf.

The reduced flux around the H$^-$ opacity minimum aligns with findings by \cite[][see their Fig.~6, right panel]{DaviesKudritzki2013}. Furthermore, the redistribution of emission from the infrared into the optical range when adopting hydrodynamical models was observed by \cite{BonifacioCaffau2018}. They noted that 3D-corrected colours are redder for hotter models and bluer for cooler models, regardless of metallicity. Their Fig.~4 demonstrates bluer $V-K$ colours from 3D models for all instances with \Teff $\ < \SI{5250}{K}$, a condition met by the models shown in Fig.~\ref{fig:sed_3d}.

\subsection{Balmer series} \label{sec:balmer}
\begin{figure*}[!ht]
    \centering
    \includegraphics[scale=\pltscale]{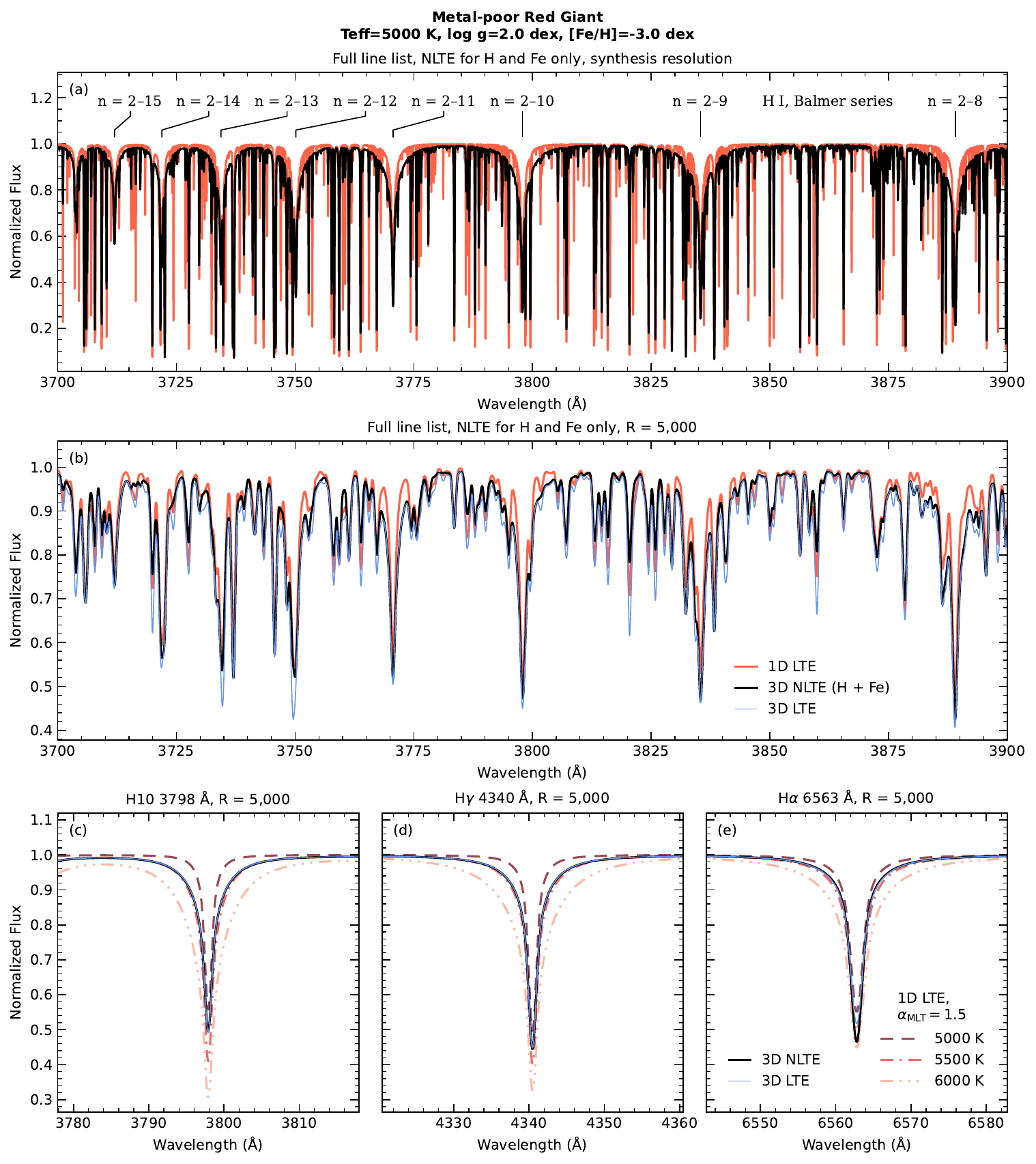}
    \caption{Synthetic spectra of a metal-poor red giant. The 3D spectra are computed from a single snapshot of an \Mdis model atmosphere, and the 1D LTE spectra come from \Marcs models. (a) The 3700--3900 \AA\ region at the synthesis resolution, computed with the line lists from Table~\ref{tab:line_lists} in 1D LTE (red) and 3D NLTE (black) where only H and Fe are treated in NLTE. The labels give the principal quantum numbers of the lower and upper levels of the Balmer lines. (b) The same spectra convolved to $R = 5000$, with 3D LTE added (blue). (c--e) H10, H$\gamma$ and H$\alpha$ at $R = 5000$, computed with hydrogen lines only. 3D NLTE (black) and 3D LTE (blue) are compared with 1D LTE on \Marcs models of 5000, 5500 and 6000 K (red).}
    \label{fig:balmer}
\end{figure*}

We present an example of a 3D NLTE spectrum synthesis, compared with the corresponding 1D LTE result, in Fig.~\ref{fig:balmer}. The 3D NLTE spectrum was computed from a single snapshot of the metal-poor red giant \Mdis model of Table~\ref{tab:stellar_parameters} in two steps. First, we computed departure coefficients for \spc{H}{1} and \spc{Fe}{1} separately, using the H and Fe model atoms of Table~\ref{tab:nlte_atoms}. We then ran a final synthesis with the line lists of Table~\ref{tab:line_lists}, applying the precomputed departure coefficients to the line opacities and emissivities of both species. This requires each line-list transition to be associated with the levels of the model atom, for which we use the optional level matching described in Sect.~\ref{sec:methods}.

Panel (a) of Fig.~\ref{fig:balmer} shows the resulting 1D LTE and 3D NLTE spectra at the synthesis resolution (sampled at $R = 200\,000$). The Balmer lines have much stronger wings in 3D NLTE than in 1D LTE. Panel (b) shows the same spectra convolved to $R = 5\,000$, together with the 3D LTE spectrum. The strong wings are equally present in 3D LTE, so they arise from the 3D atmospheric structure rather than from departures from LTE. NLTE effects are, however, visible in the many \spc{Fe}{1} lines in this window, which are weaker in 3D NLTE than in 3D LTE. In panels (c)--(e) we isolate the \Teff sensitivity of H10, H$\gamma$ and H$\alpha$ by comparing the 3D NLTE and 3D LTE profiles with 1D LTE profiles computed from \Marcs models of increasing \Teff. NLTE affects these lines mainly in the line core, and the effect decreases towards the higher series members. At $R = 5\,000$ it deepens the core of H$\alpha$ by 5\,\% of the continuum flux and changes those of H$\gamma$ and H10 by less than 0.4\,\%. Beyond about \SI{1}{\angstrom} from the centre of H$\alpha$, the 3D NLTE and 3D LTE profiles differ by less than 1\,\%. The \Marcs models were computed with a mixing length of $\alpha_{\rm MLT}$ = 1.5 which is their default value \citep{GustafssonEdvardsson2008}. At \SI{3}{\angstrom} from the line centre, the 3D NLTE wings correspond to a 1D model that is about \SI{450}{\K} hotter for H10, \SI{430}{\K} for H$\gamma$ and \SI{340}{\K} for H$\alpha$. This ordering agrees with \citet{AmarsiNordlander2018}, who found that the outer Balmer-line wings are stronger in 3D than in 1D because the line forms in the upflowing granules, whose temperature gradients are steeper than those of the 1D models, and that this effect grows towards the higher series members, which form deeper in the atmosphere. Our conclusion is that the analysis of \spc{H}{1} lines using 1D models would overestimate \Teff by up to a few hundred K (from about \SI{340}{\K} for H$\alpha$ to \SI{450}{\K} for H10 at \SI{3}{\angstrom} from the line centre) for metal-poor red giants, but the details will depend on the choice of line and fitting method.

\subsection{Time-variable spectra} \label{sec:time_series}

\begin{figure*}[!ht]
    \centering
    \includegraphics[scale=\pltscale]{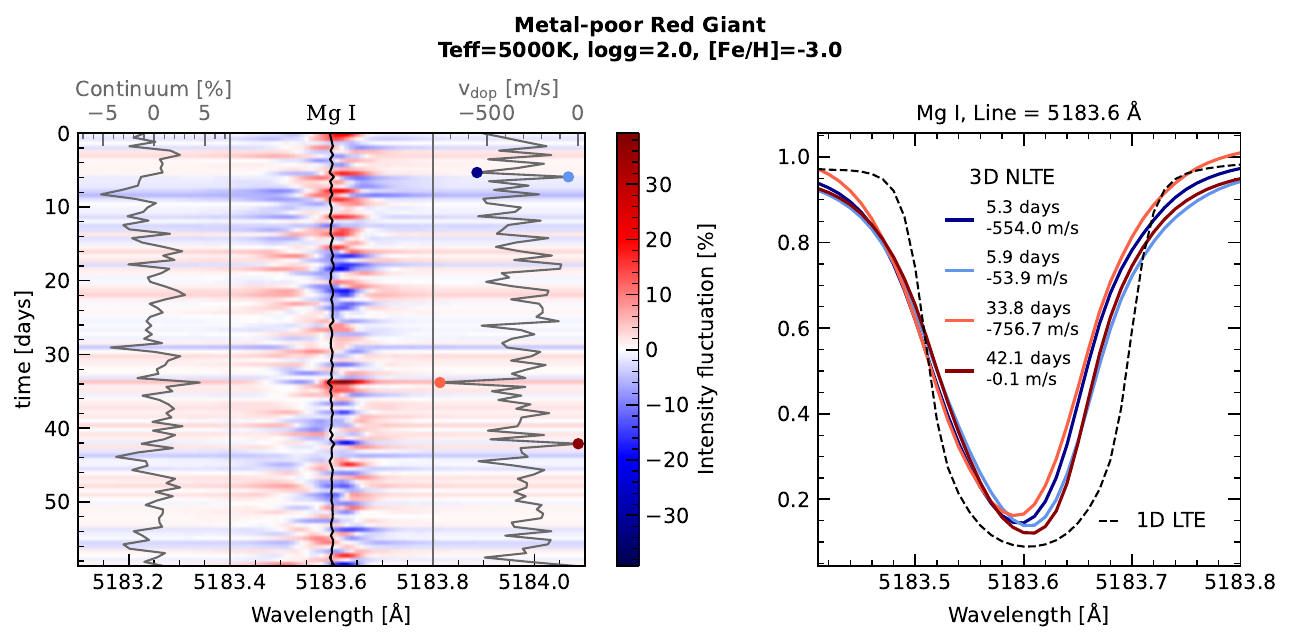}
    \caption{The left panel shows the time-variability of a 3D NLTE Mg line profile for a series of $\sim100$ snapshots, presented for a typical metal-poor red giant. The starting time is past the relaxation time (\Teff\ convergence) of the atmospheric model. The intensity fluctuations are measured with respect to the time averaged mean. The first grey inset in the left panel shows the intensity fluctuation of the continuum only. The second grey inset shows the Doppler shift of the line core, which is also shown in wavelength space by the black solid line in the centre of the plot. The right panel shows the mean line profile and highlights the line profiles from two selected snapshots responsible for the minimum and maximum Doppler shifts.}
    \label{fig:time_rgb}
\end{figure*}

\begin{figure*}[!ht]
    \centering
    \includegraphics[scale=\pltscale]{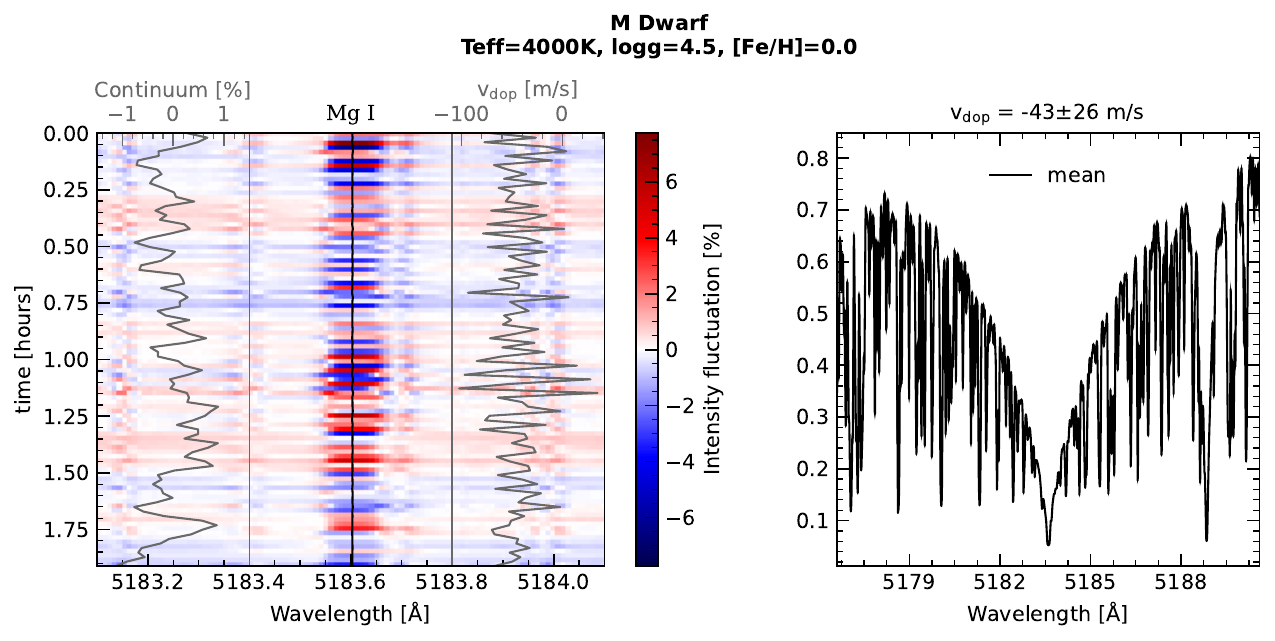}
    \caption{Same as Fig.~\ref{fig:time_rgb}, but for a typical M dwarf. We have omitted the results for individual snapshots in the right panel as the differences were hardly noticeable when zooming out to show the wings of this strong line.}
    \label{fig:time_mdw}
\end{figure*}

We performed 3D NLTE calculations of the \SI{5183.6}{\angstrom} Mg line for a sequence of snapshots representing a typical M dwarf and a metal-poor RGB star. Figs.~\ref{fig:time_rgb} and~\ref{fig:time_mdw} illustrate the intensity fluctuations surrounding and within the line core for each model. Each time series comprises approximately 100 snapshots but spans different intervals of stellar time, reflecting the disparate convective timescales of the dwarf and giant.

The intensity fluctuations were computed as:
\begin{equation}
    \Delta I_\lambda(t) = \frac{I_\lambda(t)}{\langle I_\lambda(t) \rangle} - 1,
\end{equation}

where $I_\lambda$ denotes the un-normalised disk-centre intensity and the angle brackets indicate averaging over the time series.

The intensity fluctuations comprise two categories, most clearly distinguished in Fig.~\ref{fig:time_rgb}. The first consists of fluctuations that remain nearly uniform across the line wings and the continuum, forming a horizontal barcode-like pattern. The second category corresponds to blue- or redshifts of the line core, appearing as jitter around the centre. Both phenomena arise from the balance between upward and downward flows within stellar photospheric granulation. This convective surface is composed of bright rising convection cells and dim, turbulent inter-granular lanes where cooled material sinks rapidly.

Continuum fluctuations occur because the surface area ratio between cells and lanes varies around a mean value, causing certain snapshots to appear brighter than others. The first grey inset in the left panel of Fig.~\ref{fig:time_rgb} shows the intensity fluctuations of the continuum alone. These amount to approximately 5\% for the metal-poor RGB and $\sim$1\% for the M dwarf.
\begin{figure*}[!ht]
    \centering
    \includegraphics[scale=\pltscale]{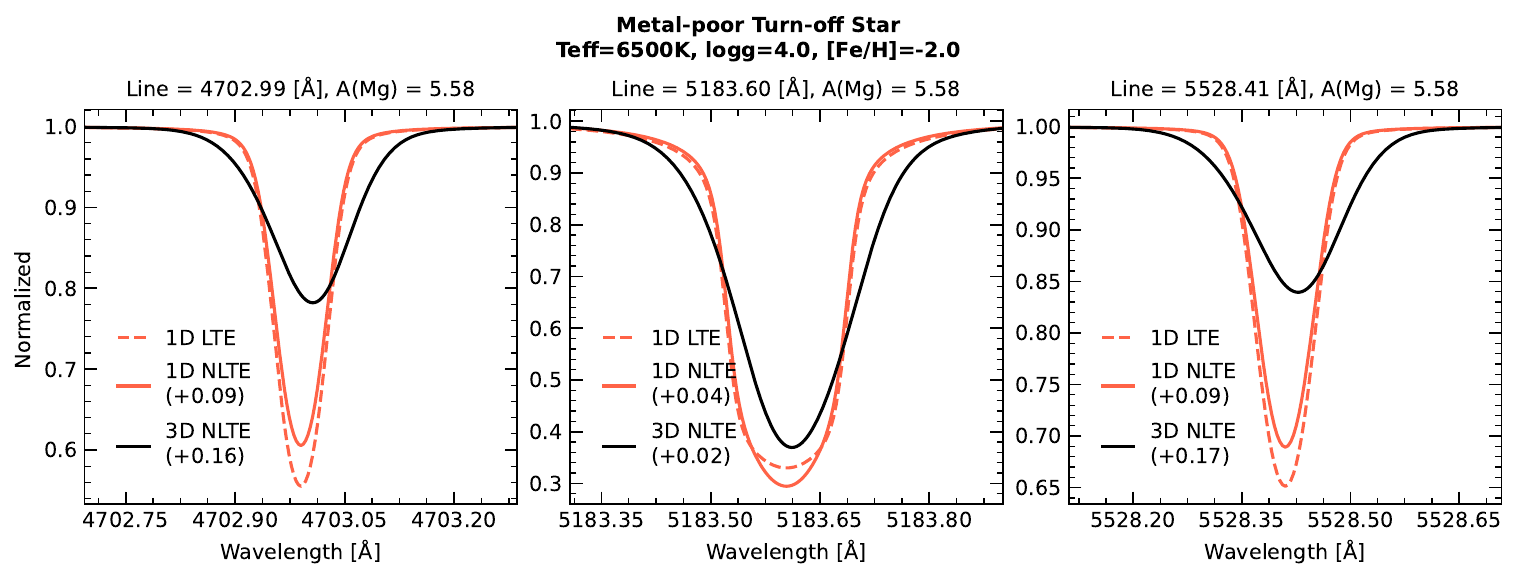}
    \caption{1D LTE, 1D NLTE and 3D NLTE \spc{Mg}{1} line profiles as well as their corresponding abundance corrections with respect to 1D LTE.}
    \label{fig:3dnlte_mg}
\end{figure*}

The convective blue- or redshifts result from a non-zero average line-of-sight momentum in the photosphere. Even when the simulation box perfectly conserves mass and momentum, balancing upward and downward flows, an imbalance typically exists in the observable fraction of the material. Consequently, the line centres appear shifted by different amounts in different snapshots \citep[e.g.][]{NordlundDravins1990, Dravins1990a}. The second grey inset in the left panel of Fig.~\ref{fig:time_rgb} shows the derived Doppler shift v$_{\rm dop}$ of the line core for each snapshot. The solid black line in the centre represents v$_{\rm dop}$ in wavelength space. The Doppler shift was measured by fitting a parabola to the line core and determining its analytical minimum. Alternatively we tested using the maximum displacement of the C-shaped line bisectors, but both methods returned consistent results in our case. The average Doppler shift velocity of the \SI{5183.6}{\angstrom} Mg line is v$_{\rm dop}$=\SI{-317(134)}{\m/\s} in the case of the metal-poor RGB and \SI{-43(26)}{\m/\s} in the case of the M dwarf. This is in good agreement with the \Teff-v$_{\rm dop}$ trend observed in FGKM-type stars predicting larger shifts for hotter stars \citep{AllendePrieto2002a, Frutiger2005, Meunier2017, Sheminova2020}. The line core intensity fluctuations peak at approximately $7\%$ for the M dwarf and $35\%$ for the metal-poor RGB star.

We note that the presented intensity fluctuations only correspond to the variability of the modelled surface area when viewed face on and it does not represent the expected variability of the integrated stellar disk. The amplitude of these fluctuations relative to the mean decreases as the horizontal extent of the simulation box increases; however, the finite stellar radius ultimately limits this reduction \citep[see e.g.][]{Ludwig2006}. The dwarf simulation box covers a horizontal area of \qtyproduct{5.08 x 5.08}{\mega\metre}, whereas the giant spans \qtyproduct{2.13 x 2.13}{\giga\metre}. We designed the horizontal dimensions of our simulation box to cover approximately 10--20 convection cells. This scale is generally considered sufficient to capture the statistical properties of the atmosphere \citep[e.g.][]{Collet2007, BergemannLind2012, MagicCollet2013, Amarsi2017}, as smaller scales typically remain spatially correlated.

\subsection{Mg abundance corrections} \label{sec:mg_corrections}
Figure~\ref{fig:3dnlte_mg} shows example calculations of 1D LTE, 1D NLTE and 3D NLTE line profiles for three \spc{Mg}{1} lines (4703.0 / 5183.6 / 5528.4) \AA\xspace in a metal-poor turn-off atmosphere (\stellarparams{6500}{4.0}{-2}). We have chosen this model atmosphere, because for this model there are four independent literature sources providing NLTE abundance correction for the \spc{Mg}{1} lines \citep{BergemannCollet2017, OsorioBarklem2016,  Mashonkina2013, LindNordlander2022}. Hence this combination of inputs (spectral lines, stellar parameters) offers an excellent validation test.

To compute the 1D and 3D NLTE abundance corrections, we used the \Marcs model atmospheres in 1D and \Stagger models from \citet{RodriguezDiazLagae2024} in the case of 3D. We employ the NLTE Mg model atom from \citet{BergemannCollet2017}, also presented in \citet{MaggBergemann2022} and \citet{GerberMagg2023}. The atomic parameters of the \spc{Mg}{1} lines are provided in Table~\ref{tab:nlte_lines}. We have also accounted for the effects of line blanketing in the background opacities through the use of a line list assembled and compiled by the Munich group \citep{GehrenShi2006, Bergemann2011, BergemannGallagher2019}. The resulting 1D NLTE $-$ 1D LTE abundance corrections are +0.09, +0.04, +0.09 dex for the \spc{Mg}{1} lines at 4703.0, 5183.6 and 5528.4 \AA, respectively. The literature results on 1D NLTE abundance corrections for these optical \spc{Mg}{1} lines are generally in good agreement (see Fig.~33 in \citet{BergemannHoppe2026}). For this set of stellar parameters, \citet{LindNordlander2022} find 1D NLTE corrections of (+0.05/+0.03/+0.04), while results based on DETAIL/SIU\footnote{\url{nlte.mpia.de}} employing the model from \citet{BergemannCollet2017} predict the corrections (+0.07/+0.03/+0.06). The NLTE abundance correction database from \citet{MatsunoAmarsi2024} tabulates 1D NLTE $-$ 1D LTE abundance corrections of (+0.06/$-$0.07/+0.04). We note that the abundance correction of the \SI{5183.6}{\angstrom} line is particularly sensitive to the inclusion of line blanketing and it can become negative in 1D NLTE if no line blanketing is accounted for. We consider the agreement satisfactory given that the three sources are employing various synthesis codes (\texttt{PySME}, \texttt{SIU} and \texttt{BALDER}) with some differences in microphysics and opacities to be expected.
For the same stellar parameters, our 3D NLTE $-$ 1D LTE corrections are (+0.16/+0.02/+0.17) dex.  For comparison, the 3D NLTE $-$ 1D LTE results from \citet{MatsunoAmarsi2024} are (+0.14/+0.06/+0.13), which agrees well with our results.

\section{Conclusions}  \label{sec:conclusions}

The standard approach to determine astrophysical parameters of stars, including atmospheric parameters and detailed chemical composition of stars, relies on highly physically simplified  synthetic stellar spectra. These spectra are typically computed using 1D hydrostatic equilibrium model atmospheres and NLTE taken into account using grids of departure coefficients, for example GALAH \citep{BuderKos2025}, GAIA-ESO \citep{SmiljanicKorn2014} and APOGEE \citep{OsorioAllendePrieto2020}. However, these models -- by construction -- do not have any time dependence and lack realistic gas motions due to convection and pulsations, the latter being ubiquitous over the entire Hertzsprung-Russell diagram \citep{Christensen-Dalsgaard2004}.

In this paper, we present the open-access code \Code to compute 3D NLTE radiation transfer and synthetic stellar spectra for AFGKM-type cool stars. The code is applicable to the following conditions: $\SI{2500}{K} \lesssim \text{\Teff}\lesssim \SI{8000}{K}$,\;$-1.0\lesssim\log g\lesssim 5.5$, and $-10\lesssim\text{[Fe/H]}\lesssim+1$, following primarily the micro-physics (partition functions, atomic and molecular line lists with a given number of ions and their line data, radiative cross-sections for continuum opacities, and temperature ranges of collisional data in the NLTE atomic and molecular models). The code can directly use 3D snapshots from 3D radiation-hydrodynamics calculations of stellar models atmospheres, such as \Cobold, \Stagger, and \Mdis. \Code is also capable of solving statistical equilibrium equations and spectrum synthesis (one or more atoms and molecules in NLTE at a time) with the departure coefficient datasets (Sect.~\ref{sec:balmer}). This code is the successor of \MultiD \citep{LeenaartsCarlsson2009}, but our code includes improved opacity calculations, updated microphysics, such as the EoS and partition functions, as well as improved interpolation and discretisation methods for solving long-characteristics radiative transfer for inclined rays. Most importantly, backward compatibility is maintained for both \Multi and \MultiD. \Code also provides interfaces for interpolating in atmosphere grids, such as the \Marcs grid, in order to provide customised synthetic spectral  calculations (for individual values of \Teff, $\log g$, [Fe/H], abundances) for realistic observed parameter space.

In this paper, we present the validation of the data products of \Code against the synthetic spectra, SEDs, and abundance corrections computed using other widely-used codes in the literature, such as \Multi \citep{Carlsson1986}, \Turb v.20 \citep{GerberMagg2023} and \MultiD \citep{LeenaartsCarlsson2009}. We show that the code provides excellent agreement with previous results, if the same input physical setup is used, e.g. the same 1D HE model atmospheres, NLTE atomic models, and stellar parameters. Also we provide some examples of the 1D NLTE and 3D NLTE abundance corrections and compare them with the literature to validate the code. We find that 3D NLTE effects are important both for modelling the stellar SEDs \citep{DaviesKudritzki2013, BonifacioCaffau2018} and for individual spectroscopic diagnostic lines of different chemical elements \citep{BergemannCollet2017, BergemannHoppe2021, MatsunoAmarsi2024, LindAmarsi2024}. We find that, for example, for metal-poor red giant model atmospheres (\stellarparams{5000}{2.0}{-3}), the 1D LTE \Marcs models may overestimate the \Teff inferred from the Balmer lines by 300--\SI{400}{\K}. This is consistent with previous values reported in \citet{LudwigBehara2009} and \citet{AmarsiNordlander2018}. We also show the predictions of the code in terms of stellar intensities computed for time-dependent series of 3D atmospheric cubes, and we briefly summarise the main features of these calculations. Specifically we find that for K-type giants (\stellarparams{5000}{2.0}{-3}) v$_{\rm dop}$ is around \SI{300}{\m\per\s}, whereas for low-mass main-sequence M dwarf (\stellarparams{4000}{4.5}{0}) the convective line shifts and asymmetries are much smaller in amplitude and are around \SI{40}{\m\per\s}. Therefore \Code enables comparison to time-dependent signals induced by planets or stellar activity, which are of similar amplitude \citep{MeunierDesort2010, MeunierLagrange2015}. However, an in depth analysis of the variability of the convective signals in the here presented simulations is beyond the scope of this paper.
\section*{Data availability}
The source code of \Code is publicly available at \url{https://github.com/Bergemann-Group-development/dispatch}.

\begin{acknowledgements}
We thank the anonymous referee for constructive comments that helped to improve the manuscript. MB, RH, and NS are supported through the Lise Meitner grant from the Max Planck Society and through the ERC grant of the European Research Council (ERC) under the European Union's Horizon 2020 research and innovation programme (Grant agreement No. 949173). RH thanks Bertrand Plez for patiently responding to numerous emails regarding \Turb and \Marcs opacities. RH implemented the \Code code and developed the Python package. MB initiated the study, provided project supervision, and contributed to tests, validation, and architecture of the \Code code and its data products. PE contributed to the code testing, validation and provided \Dispatch stellar atmosphere snapshots. NS assisted with code testing and consistency checks with \Turb through TSFitPy. AN and AP developed the \Dispatch framework and assisted with integration into the framework. JL authored the proprietary predecessor \MultiD and accepted the public release of this latest version alongside this paper. All authors contributed to the final manuscript and approved its submission. We analysed the outputs of \Code{} using NumPy, SciPy and Astropy, and generated figures using Matplotlib. Additionally, we utilised Gemini for text editing regarding readability and style and Claude Code to help identify and resolve technical inconsistencies and bottlenecks within the codebase.
\end{acknowledgements}

\bibliographystyle{aa}
\bibliography{hoppe}

\begin{thebibliography}{161}
\expandafter\ifx\csname natexlab\endcsname\relax\def\natexlab#1{#1}\fi

\bibitem[{{Ali} \& {Griem}(1966)}]{AliGriem1966}
{Ali}, A.~W. \& {Griem}, H.~R. 1966, Physical Review, 144, 366

\bibitem[{{Allende Prieto} {et~al.}(2002){Allende Prieto}, {Lambert}, {Tull},
  \& {MacQueen}}]{AllendePrieto2002a}
{Allende Prieto}, C., {Lambert}, D.~L., {Tull}, R.~G., \& {MacQueen}, P.~J.
  2002, \apjl, 566, L93

\bibitem[{{Amarsi} \& {Asplund}(2017)}]{Amarsi2017}
{Amarsi}, A.~M. \& {Asplund}, M. 2017, \mnras, 464, 264

\bibitem[{{Amarsi} {et~al.}(2019){Amarsi}, {Barklem}, {Collet}, {Grevesse}, \&
  {Asplund}}]{AmarsiBarklem2019}
{Amarsi}, A.~M., {Barklem}, P.~S., {Collet}, R., {Grevesse}, N., \& {Asplund},
  M. 2019, \aap, 624, A111

\bibitem[{{Amarsi} {et~al.}(2018){Amarsi}, {Nordlander}, {Barklem}, {Asplund},
  {Collet}, \& {Lind}}]{AmarsiNordlander2018}
{Amarsi}, A.~M., {Nordlander}, T., {Barklem}, P.~S., {et~al.} 2018, \aap, 615,
  A139

\bibitem[{{Appel} {et~al.}(2022){Appel}, {Bagdasarian}, {Basilico}, {Bellini},
  {Benziger}, {Biondi}, {Caccianiga}, {Calaprice}, {Caminata}, {Cavalcante},
  {Chepurnov}, {D'Angelo}, {Derbin}, {Di Giacinto}, {Di Marcello}, {Ding}, {Di
  Ludovico}, {Di Noto}, {Drachnev}, {Franco}, {Galbiati}, {Ghiano},
  {Giammarchi}, {Goretti}, {G{\"o}ttel}, {Gromov}, {Guffanti}, {Ianni},
  {Ianni}, {Jany}, {Kobychev}, {Korga}, {Kumaran}, {Laubenstein},
  {Litvinovich}, {Lombardi}, {Lomskaya}, {Ludhova}, {Lukyanchenko}, {Machulin},
  {Martyn}, {Meroni}, {Miramonti}, {Misiaszek}, {Muratova}, {Nugmanov},
  {Oberauer}, {Orekhov}, {Ortica}, {Pallavicini}, {Papp}, {Pelicci}, {Penek},
  {Pietrofaccia}, {Pilipenko}, {Pocar}, {Raikov}, {Ranalli}, {Ranucci},
  {Razeto}, {Re}, {Redchuk}, {Rossi}, {Sch{\"o}nert}, {Semenov}, {Settanta},
  {Skorokhvatov}, {Singhal}, {Smirnov}, {Sotnikov}, {Tartaglia}, {Testera},
  {Unzhakov}, {Villante}, {Vishneva}, {Vogelaar}, {von Feilitzsch}, {Wojcik},
  {Wurm}, {Zavatarelli}, {Zuber}, {Zuzel}, \& {Borexino
  Collaboration}}]{AppelBagdasarian2022}
{Appel}, S., {Bagdasarian}, Z., {Basilico}, D., {et~al.} 2022, \prl, 129,
  252701

\bibitem[{{Asplund} {et~al.}(2005){Asplund}, {Grevesse}, \&
  {Sauval}}]{AsplundGrevesse2005}
{Asplund}, M., {Grevesse}, N., \& {Sauval}, A.~J. 2005, in Astronomical Society
  of the Pacific Conference Series, Vol. 336, Cosmic Abundances as Records of
  Stellar Evolution and Nucleosynthesis, ed. T.~G. {Barnes}, III \& F.~N.
  {Bash}, 25

\bibitem[{{Asplund} {et~al.}(2009){Asplund}, {Grevesse}, {Sauval}, \&
  {Scott}}]{AsplundGrevesse2009}
{Asplund}, M., {Grevesse}, N., {Sauval}, A.~J., \& {Scott}, P. 2009, \araa, 47,
  481

\bibitem[{{Barber} {et~al.}(2006){Barber}, {Tennyson}, {Harris}, \&
  {Tolchenov}}]{BarberTennyson2006}
{Barber}, R.~J., {Tennyson}, J., {Harris}, G.~J., \& {Tolchenov}, R.~N. 2006,
  \mnras, 368, 1087

\bibitem[{Barklem \& Piskunov(2016)}]{BarklemPiskunov2016}
Barklem, P. \& Piskunov, N. 2016, hlinop: April 2016 release

\bibitem[{{Barklem} \& {Collet}(2016)}]{BarklemCollet2016}
{Barklem}, P.~S. \& {Collet}, R. 2016, \aap, 588, A96

\bibitem[{{Barklem} {et~al.}(2000){Barklem}, {Piskunov}, \&
  {O'Mara}}]{BarklemPiskunov2000}
{Barklem}, P.~S., {Piskunov}, N., \& {O'Mara}, B.~J. 2000, \aaps, 142, 467

\bibitem[{{Basilico} {et~al.}(2023){Basilico}, {Bellini}, {Benziger}, {Biondi},
  {Caccianiga}, {Calaprice}, {Caminata}, {Chepurnov}, {D'Angelo}, {Derbin}, {Di
  Giacinto}, {Di Marcello}, {Ding}, {Di Ludovico}, {Di Noto}, {Drachnev},
  {Franco}, {Galbiati}, {Ghiano}, {Giammarchi}, {Goretti}, {Gromov},
  {Guffanti}, {Ianni}, {Ianni}, {Jany}, {Kobychev}, {Korga}, {Kumaran},
  {Laubenstein}, {Litvinovich}, {Lombardi}, {Lomskaya}, {Ludhova}, {Machulin},
  {Martyn}, {Meroni}, {Miramonti}, {Misiaszek}, {Muratova}, {Nugmanov},
  {Oberauer}, {Orekhov}, {Ortica}, {Pallavicini}, {Pelicci}, {Penek},
  {Pietrofaccia}, {Pilipenko}, {Pocar}, {Raikov}, {Ranalli}, {Ranucci},
  {Razeto}, {Re}, {Rossi}, {Sch{\"o}nert}, {Semenov}, {Settanta},
  {Skorokhvatov}, {Singhal}, {Smirnov}, {Sotnikov}, {Tartaglia}, {Testera},
  {Unzhakov}, {Villante}, {Vishneva}, {Vogelaar}, {von Feilitzsch}, {Wojcik},
  {Wurm}, {Zavatarelli}, {Zuber}, {Zuzel}, \& {Borexino
  Collaboration}}]{BasilicoBellini2023}
{Basilico}, D., {Bellini}, G., {Benziger}, J., {et~al.} 2023, \prd, 108, 102005

\bibitem[{{Bell} \& {Berrington}(1987)}]{Bell1987}
{Bell}, K.~L. \& {Berrington}, K.~A. 1987, Journal of Physics B Atomic
  Molecular Physics, 20, 801

\bibitem[{{Belyaev} {et~al.}(2014){Belyaev}, {Yakovleva}, \&
  {Barklem}}]{BelyaevYakovleva2014}
{Belyaev}, A.~K., {Yakovleva}, S.~A., \& {Barklem}, P.~S. 2014, \aap, 572, A103

\bibitem[{{Bensby} {et~al.}(2019){Bensby}, {Bergemann}, {Rybizki}, {Lemasle},
  {Howes}, {Kovalev}, {Agertz}, {Asplund}, {Barklem}, {Battistini},
  {Casagrande}, {Chiappini}, {Church}, {Feltzing}, {Ford}, {Gerhard},
  {Kushniruk}, {Kordopatis}, {Lind}, {Minchev}, {McMillan}, {Rix}, {Ryde}, \&
  {Traven}}]{BensbyBergemann2019}
{Bensby}, T., {Bergemann}, M., {Rybizki}, J., {et~al.} 2019, The Messenger,
  175, 35

\bibitem[{{Bergemann}(2011)}]{Bergemann2011}
{Bergemann}, M. 2011, \mnras, 413, 2184

\bibitem[{{Bergemann} {et~al.}(2026{\natexlab{a}}){Bergemann}, {Bensby},
  {Gerber}, {et~al.}}]{BergemannBensby2026}
{Bergemann}, M., {Bensby}, T., {Gerber}, J., {et~al.} 2026{\natexlab{a}}, \aap,
  in press

\bibitem[{{Bergemann} \& {Cescutti}(2010)}]{BergemannCescutti2010}
{Bergemann}, M. \& {Cescutti}, G. 2010, \aap, 522, A9

\bibitem[{{Bergemann} {et~al.}(2017){Bergemann}, {Collet}, {Amarsi}, {Kovalev},
  {Ruchti}, \& {Magic}}]{BergemannCollet2017}
{Bergemann}, M., {Collet}, R., {Amarsi}, A.~M., {et~al.} 2017, \apj, 847, 15

\bibitem[{{Bergemann} {et~al.}(2019){Bergemann}, {Gallagher}, {Eitner},
  {Bautista}, {Collet}, {Yakovleva}, {Mayriedl}, {Plez}, {Carlsson},
  {Leenaarts}, {Belyaev}, \& {Hansen}}]{BergemannGallagher2019}
{Bergemann}, M., {Gallagher}, A.~J., {Eitner}, P., {et~al.} 2019, \aap, 631,
  A80

\bibitem[{{Bergemann} {et~al.}(2012{\natexlab{a}}){Bergemann}, {Hansen},
  {Bautista}, \& {Ruchti}}]{BergemannHansen2012}
{Bergemann}, M., {Hansen}, C.~J., {Bautista}, M., \& {Ruchti}, G.
  2012{\natexlab{a}}, \aap, 546, A90

\bibitem[{{Bergemann} \& {Hoppe}(2026)}]{BergemannHoppe2026}
{Bergemann}, M. \& {Hoppe}, R. 2026, Living Reviews in Computational
  Astrophysics, 12, 7

\bibitem[{{Bergemann} {et~al.}(2021){Bergemann}, {Hoppe}, {Semenova},
  {Carlsson}, {Yakovleva}, {Voronov}, {Bautista}, {Nemer}, {Belyaev},
  {Leenaarts}, {Mashonkina}, {Reiners}, \& {Ellwarth}}]{BergemannHoppe2021}
{Bergemann}, M., {Hoppe}, R., {Semenova}, E., {et~al.} 2021, \mnras, 508, 2236

\bibitem[{{Bergemann} {et~al.}(2013){Bergemann}, {Kudritzki}, {W{\"u}rl},
  {Plez}, {Davies}, \& {Gazak}}]{BergemannKudritzki2013}
{Bergemann}, M., {Kudritzki}, R.-P., {W{\"u}rl}, M., {et~al.} 2013, \apj, 764,
  115

\bibitem[{{Bergemann} {et~al.}(2012{\natexlab{b}}){Bergemann}, {Lind},
  {Collet}, {Magic}, \& {Asplund}}]{BergemannLind2012}
{Bergemann}, M., {Lind}, K., {Collet}, R., {Magic}, Z., \& {Asplund}, M.
  2012{\natexlab{b}}, \mnras, 427, 27

\bibitem[{{Bergemann} {et~al.}(2026{\natexlab{b}}){Bergemann}, {Lodders}, \&
  {Palme}}]{BergemannLodders2026}
{Bergemann}, M., {Lodders}, K., \& {Palme}, H. 2026{\natexlab{b}}, in
  Encyclopedia of Astrophysics, Volume 2, Vol.~2, 387--418

\bibitem[{{Bergemann} {et~al.}(2010){Bergemann}, {Pickering}, \&
  {Gehren}}]{BergemannPickering2010}
{Bergemann}, M., {Pickering}, J.~C., \& {Gehren}, T. 2010, \mnras, 401, 1334

\bibitem[{{Bonifacio} {et~al.}(2018){Bonifacio}, {Caffau}, {Ludwig}, {Steffen},
  {Castelli}, {Gallagher}, {Ku{\v{c}}inskas}, {Prakapavi{\v{c}}ius}, {Cayrel},
  {Freytag}, {Plez}, \& {Homeier}}]{BonifacioCaffau2018}
{Bonifacio}, P., {Caffau}, E., {Ludwig}, H.-G., {et~al.} 2018, \aap, 611, A68

\bibitem[{{Borysow} {et~al.}(2001){Borysow}, {Jorgensen}, \&
  {Fu}}]{BorysowJorgensen2001}
{Borysow}, A., {Jorgensen}, U.~G., \& {Fu}, Y. 2001, \jqsrt, 68, 235

\bibitem[{{Buder} {et~al.}(2025){Buder}, {Kos}, {Wang}, {McKenzie}, {Howell},
  {Martell}, {Hayden}, {Zucker}, {Nordlander}, {Montet}, {Traven},
  {Bland-Hawthorn}, {de Silva}, {Freeman}, {Lewis}, {Lind}, {Sharma},
  {Simpson}, {Stello}, {Zwitter}, {Amarsi}, {Armstrong}, {Banks}, {Beavis},
  {Beeson}, {Chen}, {Ciuc{\u{a}}}, {da Costa}, {de Grijs}, {Martin}, {Nataf},
  {Ness}, {Rains}, {Scarr}, {Vogrin{\v{c}}i{\v{c}}}, {Wang}, {Wittenmyer},
  {Xie}, \& {The Galah Collaboration}}]{BuderKos2025}
{Buder}, S., {Kos}, J., {Wang}, X.~E., {et~al.} 2025, \pasa, 42, e051

\bibitem[{{Bues} \& {Wehrse}(1976)}]{BuesWehrse1976}
{Bues}, I. \& {Wehrse}, R. 1976, \aap, 51, 461

\bibitem[{{Burrows} {et~al.}(2002){Burrows}, {Ram}, {Bernath}, {Sharp}, \&
  {Milsom}}]{BurrowsRam2002}
{Burrows}, A., {Ram}, R.~S., {Bernath}, P., {Sharp}, C.~M., \& {Milsom}, J.~A.
  2002, \apj, 577, 986

\bibitem[{{Cami} {et~al.}(2009){Cami}, {Sloan}, {Markwick-Kemper}, {Zijlstra},
  {Bauschlicher}, {Matsuura}, {Decin}, \& {Hony}}]{CamiSloan2009}
{Cami}, J., {Sloan}, G.~C., {Markwick-Kemper}, A.~J., {et~al.} 2009, \apjl,
  690, L122

\bibitem[{{Canocchi} {et~al.}(2024{\natexlab{a}}){Canocchi}, {Lind}, {Lagae},
  {Pietrow}, {Amarsi}, {Kiselman}, {Andriienko}, \&
  {Hoeijmakers}}]{CanocchiLind2024}
{Canocchi}, G., {Lind}, K., {Lagae}, C., {et~al.} 2024{\natexlab{a}}, \aap,
  683, A242

\bibitem[{{Canocchi} {et~al.}(2024{\natexlab{b}}){Canocchi}, {Morello}, {Lind},
  {Carleo}, {Stangret}, \& {Pall{\'e}}}]{CanocchiMorello2024}
{Canocchi}, G., {Morello}, G., {Lind}, K., {et~al.} 2024{\natexlab{b}}, \aap,
  692, A43

\bibitem[{{Carlsson}(1986)}]{Carlsson1986}
{Carlsson}, M. 1986, Uppsala Astronomical Observatory Reports, 33

\bibitem[{{Christensen-Dalsgaard}(2004)}]{Christensen-Dalsgaard2004}
{Christensen-Dalsgaard}, J. 2004, \solphys, 220, 137

\bibitem[{{Collet} {et~al.}(2007){Collet}, {Asplund}, \&
  {Trampedach}}]{Collet2007}
{Collet}, R., {Asplund}, M., \& {Trampedach}, R. 2007, \aap, 469, 687

\bibitem[{{Cowley}(1971)}]{Cowley1971}
{Cowley}, C.~R. 1971, The Observatory, 91, 139

\bibitem[{{Dalgarno} \& {Williams}(1962)}]{DalgarnoWilliams1962}
{Dalgarno}, A. \& {Williams}, D.~A. 1962, \apj, 136, 690

\bibitem[{{Davies} {et~al.}(2013){Davies}, {Kudritzki}, {Plez}, {Trager},
  {Lan{\c{c}}on}, {Gazak}, {Bergemann}, {Evans}, \&
  {Chiavassa}}]{DaviesKudritzki2013}
{Davies}, B., {Kudritzki}, R.-P., {Plez}, B., {et~al.} 2013, \apj, 767, 3

\bibitem[{{Doyle}(1968)}]{Doyle1968}
{Doyle}, R. 1968, \jqsrt, 8, 1555

\bibitem[{{Dravins}(2008)}]{Dravins2008}
{Dravins}, D. 2008, \aap, 492, 199

\bibitem[{{Dravins} \& {Ludwig}(2026)}]{DravinsLudwig2026}
{Dravins}, D. \& {Ludwig}, H.-G. 2026, in IAU Symposium, Vol. 400, Solar and
  Stellar Multi-scale Activity, ed. S.~{Vargas-Dom{\'\i}nguez} \& J.~C.
  {Buitrago-Casas} (CUP), 18--25

\bibitem[{{Dravins} {et~al.}(2017{\natexlab{a}}){Dravins}, {Ludwig},
  {Dahl{\'e}n}, \& {Pazira}}]{Dravins2017a}
{Dravins}, D., {Ludwig}, H.-G., {Dahl{\'e}n}, E., \& {Pazira}, H.
  2017{\natexlab{a}}, \aap, 605, A90

\bibitem[{{Dravins} {et~al.}(2017{\natexlab{b}}){Dravins}, {Ludwig},
  {Dahl{\'e}n}, \& {Pazira}}]{Dravins2017b}
{Dravins}, D., {Ludwig}, H.-G., {Dahl{\'e}n}, E., \& {Pazira}, H.
  2017{\natexlab{b}}, \aap, 605, A91

\bibitem[{{Dravins} {et~al.}(2021){Dravins}, {Ludwig}, \&
  {Freytag}}]{Dravins2021}
{Dravins}, D., {Ludwig}, H.-G., \& {Freytag}, B. 2021, \aap, 649, A17

\bibitem[{{Dravins} \& {Nordlund}(1990)}]{Dravins1990a}
{Dravins}, D. \& {Nordlund}, A. 1990, \aap, 228, 184

\bibitem[{{Eitner} {et~al.}(2024){Eitner}, {Bergemann}, {Hoppe}, {Nordlund},
  {Plez}, \& {Klevas}}]{EitnerBergemann2024}
{Eitner}, P., {Bergemann}, M., {Hoppe}, R., {et~al.} 2024, \aap, 688, A52

\bibitem[{{Eitner} {et~al.}(2025){Eitner}, {Bergemann}, {Hoppe}, {Storm},
  {Lipatova}, {Glover}, {Klessen}, {Nordlund}, \&
  {Popovas}}]{EitnerBergemann2025}
{Eitner}, P., {Bergemann}, M., {Hoppe}, R., {et~al.} 2025, \aap, 703, A199

\bibitem[{{Espinoza} \& {Jord{\'a}n}(2016)}]{EspinozaJordan2016}
{Espinoza}, N. \& {Jord{\'a}n}, A. 2016, \mnras, 457, 3573

\bibitem[{{Ezzeddine} {et~al.}(2018){Ezzeddine}, {Merle}, {Plez}, {Gebran},
  {Th{\'e}venin}, \& {Van der Swaelmen}}]{EzzeddineMerle2018}
{Ezzeddine}, R., {Merle}, T., {Plez}, B., {et~al.} 2018, \aap, 618, A141

\bibitem[{{Freytag} {et~al.}(2012){Freytag}, {Steffen}, {Ludwig},
  {Wedemeyer-B{\"o}hm}, {Schaffenberger}, \& {Steiner}}]{Freytag2012}
{Freytag}, B., {Steffen}, M., {Ludwig}, H.~G., {et~al.} 2012, Journal of
  Computational Physics, 231, 919

\bibitem[{{Frutiger} {et~al.}(2005){Frutiger}, {Solanki}, \&
  {Mathys}}]{Frutiger2005}
{Frutiger}, C., {Solanki}, S.~K., \& {Mathys}, G. 2005, \aap, 444, 549

\bibitem[{{Gallagher} {et~al.}(2020){Gallagher}, {Bergemann}, {Collet}, {Plez},
  {Leenaarts}, {Carlsson}, {Yakovleva}, \& {Belyaev}}]{GallagherBergemann2020}
{Gallagher}, A.~J., {Bergemann}, M., {Collet}, R., {et~al.} 2020, \aap, 634,
  A55

\bibitem[{{Gehren} {et~al.}(2006){Gehren}, {Shi}, {Zhang}, {Zhao}, \&
  {Korn}}]{GehrenShi2006}
{Gehren}, T., {Shi}, J.~R., {Zhang}, H.~W., {Zhao}, G., \& {Korn}, A.~J. 2006,
  \aap, 451, 1065

\bibitem[{{Gerber} {et~al.}(2023){Gerber}, {Magg}, {Plez}, {Bergemann},
  {Heiter}, {Olander}, \& {Hoppe}}]{GerberMagg2023}
{Gerber}, J.~M., {Magg}, E., {Plez}, B., {et~al.} 2023, \aap, 669, A43

\bibitem[{{Gingerich}(1964)}]{Gingerich1964}
{Gingerich}, O. 1964, SAO Special Report, 167, 17

\bibitem[{{Goldman} {et~al.}(1998){Goldman}, {Schoenfeld}, {Goorvitch},
  {Chackerian}, {Dothe}, {M{\'e}len}, {Abrams}, \&
  {Selby}}]{GoldmanSchoenfeld1998}
{Goldman}, A., {Schoenfeld}, W.~G., {Goorvitch}, D., {et~al.} 1998, \jqsrt, 59,
  453

\bibitem[{{Gonzalez-Garcia} {et~al.}(2024){Gonzalez-Garcia}, {Maltoni},
  {Pinheiro}, \& {Serenelli}}]{Gonzalez-GarciaMaltoni2024}
{Gonzalez-Garcia}, M.~C., {Maltoni}, M., {Pinheiro}, J.~P., \& {Serenelli},
  A.~M. 2024, Journal of High Energy Physics, 2024, 64

\bibitem[{{Goorvitch}(1994)}]{Goorvitch1994}
{Goorvitch}, D. 1994, \apjs, 95, 535

\bibitem[{{Gordon} {et~al.}(2017){Gordon}, {Rothman}, {Hill}, {Kochanov},
  {Tan}, {Bernath}, {Birk}, {Boudon}, {Campargue}, {Chance}, {Drouin}, {Flaud},
  {Gamache}, {Hodges}, {Jacquemart}, {Perevalov}, {Perrin}, {Shine}, {Smith},
  {Tennyson}, {Toon}, {Tran}, {Tyuterev}, {Barbe}, {Cs{\'a}sz{\'a}r}, {Devi},
  {Furtenbacher}, {Harrison}, {Hartmann}, {Jolly}, {Johnson}, {Karman},
  {Kleiner}, {Kyuberis}, {Loos}, {Lyulin}, {Massie}, {Mikhailenko},
  {Moazzen-Ahmadi}, {M{\"u}ller}, {Naumenko}, {Nikitin}, {Polyansky}, {Rey},
  {Rotger}, {Sharpe}, {Sung}, {Starikova}, {Tashkun}, {Auwera}, {Wagner},
  {Wilzewski}, {Wcis{\l}o}, {Yu}, \& {Zak}}]{GordonRothman2017}
{Gordon}, I.~E., {Rothman}, L.~S., {Hill}, C., {et~al.} 2017, \jqsrt, 203, 3

\bibitem[{{Gray}(2005)}]{Gray2005}
{Gray}, D.~F. 2005, {The Observation and Analysis of Stellar Photospheres}, 3rd
  edn. (Cambridge, UK: Cambridge University Press)

\bibitem[{{Grevesse} \& {Sauval}(1998)}]{GrevesseSauval1998}
{Grevesse}, N. \& {Sauval}, A.~J. 1998, \ssr, 85, 161

\bibitem[{{Grupp}(2004{\natexlab{a}})}]{Grupp2004a}
{Grupp}, F. 2004{\natexlab{a}}, \aap, 420, 289

\bibitem[{{Grupp}(2004{\natexlab{b}})}]{Grupp2004b}
{Grupp}, F. 2004{\natexlab{b}}, \aap, 426, 309

\bibitem[{{Gustafsson} {et~al.}(1975){Gustafsson}, {Bell}, {Eriksson}, \&
  {Nordlund}}]{GustafssonBell1975}
{Gustafsson}, B., {Bell}, R.~A., {Eriksson}, K., \& {Nordlund}, A. 1975, \aap,
  42, 407

\bibitem[{{Gustafsson} {et~al.}(2008){Gustafsson}, {Edvardsson}, {Eriksson},
  {J{\o}rgensen}, {Nordlund}, \& {Plez}}]{GustafssonEdvardsson2008}
{Gustafsson}, B., {Edvardsson}, B., {Eriksson}, K., {et~al.} 2008, \aap, 486,
  951

\bibitem[{{Gustafsson} \& {Frommhold}(2001)}]{GustafssonFrommhold2001}
{Gustafsson}, M. \& {Frommhold}, L. 2001, \apj, 546, 1168

\bibitem[{{Gustafsson} \& {Frommhold}(2003)}]{GustafssonFrommhold2003}
{Gustafsson}, M. \& {Frommhold}, L. 2003, \aap, 400, 1161

\bibitem[{{Hauschildt} \& {Baron}(2014)}]{HauschildtBaron2014}
{Hauschildt}, P.~H. \& {Baron}, E. 2014, \aap, 566, A89

\bibitem[{{Hayek} {et~al.}(2011){Hayek}, {Asplund}, {Collet}, \&
  {Nordlund}}]{HayekAsplund2011}
{Hayek}, W., {Asplund}, M., {Collet}, R., \& {Nordlund}, {\r{A}}. 2011, \aap,
  529, A158

\bibitem[{{Hedrosa} {et~al.}(2013){Hedrosa}, {Abia}, {Busso}, {Cristallo},
  {Dom{\'\i}nguez}, {Palmerini}, {Plez}, \& {Straniero}}]{HedrosaAbia2013}
{Hedrosa}, R.~P., {Abia}, C., {Busso}, M., {et~al.} 2013, \apjl, 768, L11

\bibitem[{{Hoppe} {et~al.}(2026){Hoppe}, {Bergemann}, {Eitner}, {Ellwarth},
  {Nordlund}, {Leenaarts}, {Plez}, \& {Serenelli}}]{HoppeBergemann2026}
{Hoppe}, R., {Bergemann}, M., {Eitner}, P., {et~al.} 2026, \mnras, 546,
  staf2085

\bibitem[{{Hubeny} {et~al.}(2021){Hubeny}, {Allende Prieto}, {Osorio}, \&
  {Lanz}}]{HubenyAllendePrieto2021}
{Hubeny}, I., {Allende Prieto}, C., {Osorio}, Y., \& {Lanz}, T. 2021, arXiv
  e-prints, arXiv:2104.02829

\bibitem[{{Hummer} \& {Mihalas}(1988)}]{HummerMihalas1988}
{Hummer}, D.~G. \& {Mihalas}, D. 1988, \apj, 331, 794

\bibitem[{{Irwin}(1981)}]{Irwin1981}
{Irwin}, A.~W. 1981, \apjs, 45, 621

\bibitem[{{Irwin}(1988)}]{Irwin1988}
{Irwin}, A.~W. 1988, \aaps, 74, 145

\bibitem[{{Jofr{\'e}} {et~al.}(2019){Jofr{\'e}}, {Heiter}, \&
  {Soubiran}}]{JofreHeiter2019}
{Jofr{\'e}}, P., {Heiter}, U., \& {Soubiran}, C. 2019, \araa, 57, 571

\bibitem[{{John}(1975{\natexlab{a}})}]{John1975b}
{John}, T.~L. 1975{\natexlab{a}}, \mnras, 172, 305

\bibitem[{{John}(1975{\natexlab{b}})}]{John1975a}
{John}, T.~L. 1975{\natexlab{b}}, \mnras, 170, 5

\bibitem[{{John}(1994)}]{John1994}
{John}, T.~L. 1994, \mnras, 269, 871

\bibitem[{{Jones}(1973)}]{Jones1973}
{Jones}, H.~P. 1973, \apj, 185, 183

\bibitem[{{Jones} \& {Skumanich}(1973)}]{JonesSkumanich1973}
{Jones}, H.~P. \& {Skumanich}, A. 1973, \apj, 185, 167

\bibitem[{{J{\o}rgensen} {et~al.}(2000){J{\o}rgensen}, {Hammer}, {Borysow}, \&
  {Falkesgaard}}]{JorgensenHammer2000}
{J{\o}rgensen}, U.~G., {Hammer}, D., {Borysow}, A., \& {Falkesgaard}, J. 2000,
  \aap, 361, 283

\bibitem[{{Karzas} \& {Latter}(1961)}]{Karzas1961}
{Karzas}, W.~J. \& {Latter}, R. 1961, \apjs, 6, 167

\bibitem[{{Klevas} {et~al.}(2016){Klevas}, {Ku{\v{c}}inskas}, {Steffen},
  {Caffau}, \& {Ludwig}}]{KlevasKucinskas2016}
{Klevas}, J., {Ku{\v{c}}inskas}, A., {Steffen}, M., {Caffau}, E., \& {Ludwig},
  H.-G. 2016, \aap, 586, A156

\bibitem[{{Korn} {et~al.}(2003){Korn}, {Shi}, \& {Gehren}}]{KornShi2003}
{Korn}, A.~J., {Shi}, J., \& {Gehren}, T. 2003, \aap, 407, 691

\bibitem[{{Kunasz} \& {Auer}(1988)}]{Kunasz1988}
{Kunasz}, P. \& {Auer}, L.~H. 1988, \jqsrt, 39, 67

\bibitem[{{Kupka} {et~al.}(2000){Kupka}, {Ryabchikova}, {Piskunov}, {Stempels},
  \& {Weiss}}]{KupkaRyabchikova2000}
{Kupka}, F.~G., {Ryabchikova}, T.~A., {Piskunov}, N.~E., {Stempels}, H.~C., \&
  {Weiss}, W.~W. 2000, Baltic Astronomy, 9, 590

\bibitem[{{Kurucz}(1970)}]{Kurucz1970}
{Kurucz}, R.~L. 1970, SAO Special Report, 309

\bibitem[{{Kurucz}(2011)}]{Kurucz2011}
{Kurucz}, R.~L. 2011, Canadian Journal of Physics, 89, 417

\bibitem[{{Kurucz} {et~al.}(1987){Kurucz}, {van Dishoeck}, \&
  {Tarafdar}}]{Kurucz1987}
{Kurucz}, R.~L., {van Dishoeck}, E.~F., \& {Tarafdar}, S.~P. 1987, \apj, 322,
  992

\bibitem[{{Langhoff} \& {Bauschlicher}(1993)}]{LanghoffBauschlicher1993}
{Langhoff}, S.~R. \& {Bauschlicher}, Jr., C.~W. 1993, Chemical Physics Letters,
  211, 305

\bibitem[{{Leenaarts} \& {Carlsson}(2009)}]{LeenaartsCarlsson2009}
{Leenaarts}, J. \& {Carlsson}, M. 2009, in Astronomical Society of the Pacific
  Conference Series, Vol. 415, The Second Hinode Science Meeting: Beyond
  Discovery-Toward Understanding, ed. B.~{Lites}, M.~{Cheung}, T.~{Magara},
  J.~{Mariska}, \& K.~{Reeves}, 87

\bibitem[{{Lemke}(1997)}]{Lemke1997}
{Lemke}, M. 1997, \aaps, 122, 285

\bibitem[{{Li} {et~al.}(2015){Li}, {Gordon}, {Rothman}, {Tan}, {Hu}, {Kassi},
  {Campargue}, \& {Medvedev}}]{LiGordon2015}
{Li}, G., {Gordon}, I.~E., {Rothman}, L.~S., {et~al.} 2015, \apjs, 216, 15

\bibitem[{{Lind} \& {Amarsi}(2024)}]{LindAmarsi2024}
{Lind}, K. \& {Amarsi}, A.~M. 2024, \araa, 62, 475

\bibitem[{{Lind} {et~al.}(2022){Lind}, {Nordlander}, {Wehrhahn}, {Montelius},
  {Osorio}, {Barklem}, {Af{\c{s}}ar}, {Sneden}, \&
  {Kobayashi}}]{LindNordlander2022}
{Lind}, K., {Nordlander}, T., {Wehrhahn}, A., {et~al.} 2022, \aap, 665, A33

\bibitem[{{Ludwig}(2006)}]{Ludwig2006}
{Ludwig}, H.-G. 2006, \aap, 445, 661

\bibitem[{{Ludwig} {et~al.}(2009){Ludwig}, {Behara}, {Steffen}, \&
  {Bonifacio}}]{LudwigBehara2009}
{Ludwig}, H.-G., {Behara}, N.~T., {Steffen}, M., \& {Bonifacio}, P. 2009, \aap,
  502, L1

\bibitem[{{Magg} {et~al.}(2022){Magg}, {Bergemann}, {Serenelli}, {Bautista},
  {Plez}, {Heiter}, {Gerber}, {Ludwig}, {Basu}, {Ferguson}, {Gallego},
  {Gamrath}, {Palmeri}, \& {Quinet}}]{MaggBergemann2022}
{Magg}, E., {Bergemann}, M., {Serenelli}, A., {et~al.} 2022, \aap, 661, A140

\bibitem[{{Magic} {et~al.}(2013){Magic}, {Collet}, {Asplund}, {Trampedach},
  {Hayek}, {Chiavassa}, {Stein}, \& {Nordlund}}]{MagicCollet2013}
{Magic}, Z., {Collet}, R., {Asplund}, M., {et~al.} 2013, \aap, 557, A26

\bibitem[{{Mashonkina}(2013)}]{Mashonkina2013}
{Mashonkina}, L. 2013, \aap, 550, A28

\bibitem[{{Mashonkina} {et~al.}(2007){Mashonkina}, {Korn}, \&
  {Przybilla}}]{MashonkinaKorn2007}
{Mashonkina}, L., {Korn}, A.~J., \& {Przybilla}, N. 2007, \aap, 461, 261

\bibitem[{{Mashonkina} {et~al.}(2008){Mashonkina}, {Zhao}, {Gehren}, {Aoki},
  {Bergemann}, {Noguchi}, {Shi}, {Takada-Hidai}, \&
  {Zhang}}]{MashonkinaZhao2008}
{Mashonkina}, L., {Zhao}, G., {Gehren}, T., {et~al.} 2008, \aap, 478, 529

\bibitem[{{Masseron} {et~al.}(2014){Masseron}, {Plez}, {Van Eck}, {Colin},
  {Daoutidis}, {Godefroid}, {Coheur}, {Bernath}, {Jorissen}, \&
  {Christlieb}}]{MasseronPlez2014}
{Masseron}, T., {Plez}, B., {Van Eck}, S., {et~al.} 2014, \aap, 571, A47

\bibitem[{{Matsuno} {et~al.}(2024){Matsuno}, {Amarsi}, {Carlos}, \&
  {Nissen}}]{MatsunoAmarsi2024}
{Matsuno}, T., {Amarsi}, A.~M., {Carlos}, M., \& {Nissen}, P.~E. 2024, \aap,
  688, A72

\bibitem[{{Maxted}(2018)}]{Maxted2018}
{Maxted}, P.~F.~L. 2018, \aap, 616, A39

\bibitem[{{McLaughlin} {et~al.}(2017){McLaughlin}, {Stancil}, {Sadeghpour}, \&
  {Forrey}}]{McLaughlin2017}
{McLaughlin}, B.~M., {Stancil}, P.~C., {Sadeghpour}, H.~R., \& {Forrey}, R.~C.
  2017, Journal of Physics B Atomic Molecular Physics, 50, 114001

\bibitem[{{M{\'e}sz{\'a}ros} {et~al.}(2012){M{\'e}sz{\'a}ros}, {Allende
  Prieto}, {Edvardsson}, {Castelli}, {Garc{\'\i}a P{\'e}rez}, {Gustafsson},
  {Majewski}, {Plez}, {Schiavon}, {Shetrone}, \& {de
  Vicente}}]{MeszarosAllendePrieto2012}
{M{\'e}sz{\'a}ros}, S., {Allende Prieto}, C., {Edvardsson}, B., {et~al.} 2012,
  \aj, 144, 120

\bibitem[{{Meunier} {et~al.}(2010){Meunier}, {Desort}, \&
  {Lagrange}}]{MeunierDesort2010}
{Meunier}, N., {Desort}, M., \& {Lagrange}, A.-M. 2010, \aap, 512, A39

\bibitem[{{Meunier} {et~al.}(2015){Meunier}, {Lagrange}, {Borgniet}, \&
  {Rieutord}}]{MeunierLagrange2015}
{Meunier}, N., {Lagrange}, A.-M., {Borgniet}, S., \& {Rieutord}, M. 2015, \aap,
  583, A118

\bibitem[{{Meunier} {et~al.}(2017){Meunier}, {Mignon}, \&
  {Lagrange}}]{Meunier2017}
{Meunier}, N., {Mignon}, L., \& {Lagrange}, A.~M. 2017, \aap, 607, A124

\bibitem[{{Mihalas}(1965)}]{Mihalas1965}
{Mihalas}, D. 1965, \apjs, 9, 321

\bibitem[{{Mihalas}(1978)}]{Mihalas1978}
{Mihalas}, D. 1978, {Stellar atmospheres}, 2nd edn. (San Francisco: W. H.
  Freeman)

\bibitem[{{Mihalas} {et~al.}(1978){Mihalas}, {Auer}, \&
  {Mihalas}}]{MihalasAuer1978}
{Mihalas}, D., {Auer}, L.~H., \& {Mihalas}, B.~R. 1978, \apj, 220, 1001

\bibitem[{{Ng}(1974)}]{Ng1974}
{Ng}, K.-C. 1974, \jcp, 61, 2680

\bibitem[{{Nissen} \& {Gustafsson}(2018)}]{NissenGustafsson2018}
{Nissen}, P.~E. \& {Gustafsson}, B. 2018, \aapr, 26, 6

\bibitem[{{Nordlund} \& {Dravins}(1990)}]{NordlundDravins1990}
{Nordlund}, A. \& {Dravins}, D. 1990, \aap, 228, 155

\bibitem[{{Nordlund} {et~al.}(2018){Nordlund}, {Ramsey}, {Popovas}, \&
  {K{\"u}ffmeier}}]{NordlundRamsey2018}
{Nordlund}, {\r{A}}., {Ramsey}, J.~P., {Popovas}, A., \& {K{\"u}ffmeier}, M.
  2018, \mnras, 477, 624

\bibitem[{{Ohta} {et~al.}(2005){Ohta}, {Taruya}, \& {Suto}}]{OhtaTaruya2005}
{Ohta}, Y., {Taruya}, A., \& {Suto}, Y. 2005, \apj, 622, 1118

\bibitem[{{Osorio} {et~al.}(2020){Osorio}, {Allende Prieto}, {Hubeny},
  {M{\'e}sz{\'a}ros}, \& {Shetrone}}]{OsorioAllendePrieto2020}
{Osorio}, Y., {Allende Prieto}, C., {Hubeny}, I., {M{\'e}sz{\'a}ros}, S., \&
  {Shetrone}, M. 2020, \aap, 637, A80

\bibitem[{{Osorio} \& {Barklem}(2016)}]{OsorioBarklem2016}
{Osorio}, Y. \& {Barklem}, P.~S. 2016, \aap, 586, A120

\bibitem[{{Peach}(1970)}]{Peach1970}
{Peach}, G. 1970, \memras, 73, 1

\bibitem[{{Pecaut} \& {Mamajek}(2013)}]{PecautMamajek2013}
{Pecaut}, M.~J. \& {Mamajek}, E.~E. 2013, \apjs, 208, 9

\bibitem[{{Peck} {et~al.}(2017){Peck}, {Criscuoli}, \&
  {Rast}}]{PeckCriscuoli2017}
{Peck}, C.~L., {Criscuoli}, S., \& {Rast}, M.~P. 2017, \apj, 850, 9

\bibitem[{{Piskunov} {et~al.}(1995){Piskunov}, {Kupka}, {Ryabchikova}, {Weiss},
  \& {Jeffery}}]{PiskunovKupka1995}
{Piskunov}, N.~E., {Kupka}, F., {Ryabchikova}, T.~A., {Weiss}, W.~W., \&
  {Jeffery}, C.~S. 1995, \aaps, 112, 525

\bibitem[{{Plez}(1998)}]{Plez1998}
{Plez}, B. 1998, \aap, 337, 495

\bibitem[{{Plez}(2012)}]{Plez2012}
{Plez}, B. 2012, {Turbospectrum: Code for spectral synthesis}, Astrophysics
  Source Code Library, record ascl:1205.004

\bibitem[{{Querci} {et~al.}(1971){Querci}, {Querci}, \&
  {Kunde}}]{QuerciQuerci1971}
{Querci}, F., {Querci}, M., \& {Kunde}, V.~G. 1971, \aap, 15, 256

\bibitem[{{Ramsbottom} {et~al.}(1992){Ramsbottom}, {Bell}, \&
  {Berrington}}]{RamsbottomBell1992}
{Ramsbottom}, C.~A., {Bell}, K.~L., \& {Berrington}, K.~A. 1992, Journal of
  Physics B Atomic Molecular Physics, 25, 1443

\bibitem[{{Reetz}(1991)}]{Reetz1991}
{Reetz}, J.~K. 1991, Diploma thesis, Ludwig-Maximilians-Universit{\"a}t
  M{\"u}nchen

\bibitem[{{Rodr{\'\i}guez D{\'\i}az} {et~al.}(2024){Rodr{\'\i}guez D{\'\i}az},
  {Lagae}, {Amarsi}, {Bigot}, {Zhou}, {Aguirre B{\o}rsen-Koch}, {Lind},
  {Trampedach}, \& {Collet}}]{RodriguezDiazLagae2024}
{Rodr{\'\i}guez D{\'\i}az}, L.~F., {Lagae}, C., {Amarsi}, A.~M., {et~al.} 2024,
  \aap, 688, A212

\bibitem[{{R{\'o}{\.z}a{\'n}ski} {et~al.}(2025){R{\'o}{\.z}a{\'n}ski}, {Ting},
  \& {Jab{\l}o{\'n}ska}}]{RozanskiTing2025}
{R{\'o}{\.z}a{\'n}ski}, T., {Ting}, Y.-S., \& {Jab{\l}o{\'n}ska}, M. 2025,
  \apj, 980, 66

\bibitem[{{Rybicki} \& {Hummer}(1992)}]{RybickiHummer1992}
{Rybicki}, G.~B. \& {Hummer}, D.~G. 1992, \aap, 262, 209

\bibitem[{{Sch{\"o}nrich} \& {Bergemann}(2014)}]{SchonrichBergemann2014}
{Sch{\"o}nrich}, R. \& {Bergemann}, M. 2014, \mnras, 443, 698

\bibitem[{{Seaton} {et~al.}(1994){Seaton}, {Yan}, {Mihalas}, \&
  {Pradhan}}]{Seaton1994}
{Seaton}, M.~J., {Yan}, Y., {Mihalas}, D., \& {Pradhan}, A.~K. 1994, \mnras,
  266, 805

\bibitem[{{Semenova} {et~al.}(2020){Semenova}, {Bergemann}, {Deal},
  {Serenelli}, {Hansen}, {Gallagher}, {Bayo}, {Bensby}, {Bragaglia}, {Carraro},
  {Morbidelli}, {Pancino}, \& {Smiljanic}}]{SemenovaBergemann2020}
{Semenova}, E., {Bergemann}, M., {Deal}, M., {et~al.} 2020, \aap, 643, A164

\bibitem[{{Serenelli} {et~al.}(2009){Serenelli}, {Basu}, {Ferguson}, \&
  {Asplund}}]{SerenelliBasu2009}
{Serenelli}, A.~M., {Basu}, S., {Ferguson}, J.~W., \& {Asplund}, M. 2009,
  \apjl, 705, L123

\bibitem[{{Sheminova}(2020)}]{Sheminova2020}
{Sheminova}, V.~A. 2020, Kinematics and Physics of Celestial Bodies, 36, 291

\bibitem[{{Skory} {et~al.}(2003){Skory}, {Weck}, {Stancil}, \&
  {Kirby}}]{SkoryWeck2003}
{Skory}, S., {Weck}, P.~F., {Stancil}, P.~C., \& {Kirby}, K. 2003, \apjs, 148,
  599

\bibitem[{{Smiljanic} {et~al.}(2014){Smiljanic}, {Korn}, {Bergemann}, {Frasca},
  {Magrini}, {Masseron}, {Pancino}, {Ruchti}, {San Roman}, {Sbordone}, {Sousa},
  {Tabernero}, {Tautvai{\v{s}}ien{\.{e}}}, {Valentini}, {Weber}, {Worley},
  {Adibekyan}, {Allende Prieto}, {Barisevi{\v{c}}ius}, {Biazzo},
  {Blanco-Cuaresma}, {Bonifacio}, {Bragaglia}, {Caffau}, {Cantat-Gaudin},
  {Chorniy}, {de Laverny}, {Delgado-Mena}, {Donati}, {Duffau}, {Franciosini},
  {Friel}, {Geisler}, {Gonz{\'a}lez Hern{\'a}ndez}, {Gruyters}, {Guiglion},
  {Hansen}, {Heiter}, {Hill}, {Jacobson}, {Jofre}, {J{\"o}nsson}, {Lanzafame},
  {Lardo}, {Ludwig}, {Maiorca}, {Mikolaitis}, {Montes}, {Morel}, {Mucciarelli},
  {Mu{\~n}oz}, {Nordlander}, {Pasquini}, {Puzeras}, {Recio-Blanco}, {Ryde},
  {Sacco}, {Santos}, {Serenelli}, {Sordo}, {Soubiran}, {Spina}, {Steffen},
  {Vallenari}, {Van Eck}, {Villanova}, {Gilmore}, {Randich}, {Asplund},
  {Binney}, {Drew}, {Feltzing}, {Ferguson}, {Jeffries}, {Micela}, {Negueruela},
  {Prusti}, {Rix}, {Alfaro}, {Babusiaux}, {Bensby}, {Blomme}, {Flaccomio},
  {Fran{\c{c}}ois}, {Irwin}, {Koposov}, {Walton}, {Bayo}, {Carraro}, {Costado},
  {Damiani}, {Edvardsson}, {Hourihane}, {Jackson}, {Lewis}, {Lind}, {Marconi},
  {Martayan}, {Monaco}, {Morbidelli}, {Prisinzano}, \&
  {Zaggia}}]{SmiljanicKorn2014}
{Smiljanic}, R., {Korn}, A.~J., {Bergemann}, M., {et~al.} 2014, \aap, 570, A122

\bibitem[{{Steffen} {et~al.}(2015){Steffen}, {Prakapavi{\v{c}}ius}, {Caffau},
  {Ludwig}, {Bonifacio}, {Cayrel}, {Ku{\v{c}}inskas}, \&
  {Livingston}}]{SteffenPrakapavicius2015}
{Steffen}, M., {Prakapavi{\v{c}}ius}, D., {Caffau}, E., {et~al.} 2015, \aap,
  583, A57

\bibitem[{{Stehl{\'e}} \& {Hutcheon}(1999)}]{StehleHutcheon1999}
{Stehl{\'e}}, C. \& {Hutcheon}, R. 1999, \aaps, 140, 93

\bibitem[{{Stein} {et~al.}(2024){Stein}, {Nordlund}, {Collet}, \&
  {Trampedach}}]{SteinNordlund2024}
{Stein}, R.~F., {Nordlund}, {\r{A}}., {Collet}, R., \& {Trampedach}, R. 2024,
  \apj, 970, 24

\bibitem[{{Storm} {et~al.}(2024){Storm}, {Barklem}, {Yakovleva}, {Belyaev},
  {Palmeri}, {Quinet}, {Lodders}, {Bergemann}, \& {Hoppe}}]{StormBarklem2024}
{Storm}, N., {Barklem}, P.~S., {Yakovleva}, S.~A., {et~al.} 2024, \aap, 683,
  A200

\bibitem[{{Storm} \& {Bergemann}(2023)}]{StormBergemann2023}
{Storm}, N. \& {Bergemann}, M. 2023, \mnras, 525, 3718

\bibitem[{{Storm} {et~al.}(2025){Storm}, {Bergemann}, {Eitner}, {Hoppe},
  {Kemp}, {Ruiter}, {Janka}, {Sieverding}, {de Mink}, {Seitenzahl}, \&
  et~al.}]{StormBergemann2025}
{Storm}, N., {Bergemann}, M., {Eitner}, P., {et~al.} 2025, \mnras, 538, 3284

\bibitem[{{Storm} {et~al.}(2026){Storm}, {Bergemann}, {R{\'o}{\.z}a{\'n}ski},
  {F. Ksoll}, {Bensby}, {Traven}, {Kordopatis}, {Church}, {Jian}, {Sun}, \&
  et~al.}]{StormBergemann2026}
{Storm}, N., {Bergemann}, M., {R{\'o}{\.z}a{\'n}ski}, T., {et~al.} 2026, \apj,
  1003, 27

\bibitem[{{Tennyson} {et~al.}(2024){Tennyson}, {Yurchenko}, {Zhang},
  {Bowesman}, {Brady}, {Buldyreva}, {Chubb}, {Gamache}, {Gorman}, {Guest},
  {Hill}, {Kefala}, {Lynas-Gray}, {Mellor}, {McKemmish}, {Mitev}, {Mizus},
  {Owens}, {Peng}, {Perri}, {Pezzella}, {Polyansky}, {Qu}, {Semenov}, {Smola},
  {Solokov}, {Somogyi}, {Upadhyay}, {Wright}, \&
  {Zobov}}]{TennysonYurchenko2024}
{Tennyson}, J., {Yurchenko}, S.~N., {Zhang}, J., {et~al.} 2024, \jqsrt, 326,
  109083

\bibitem[{{Uns\"old}(1955)}]{Unsold1955}
{Uns\"old}, A. 1955, {Physik der Sternatmosph\"aren, mit besonderer
  Ber\"ucksichtigung der Sonne}, 2nd edn. (Berlin: Springer)

\bibitem[{{Vidal} {et~al.}(1973){Vidal}, {Cooper}, \&
  {Smith}}]{VidalCooper1973}
{Vidal}, C.~R., {Cooper}, J., \& {Smith}, E.~W. 1973, \apjs, 25, 37

\bibitem[{{V{\"o}gler} {et~al.}(2004){V{\"o}gler}, {Bruls}, \&
  {Sch{\"u}ssler}}]{Vogler2004}
{V{\"o}gler}, A., {Bruls}, J.~H.~M.~J., \& {Sch{\"u}ssler}, M. 2004, \aap, 421,
  741

\bibitem[{{V{\"o}gler} {et~al.}(2005){V{\"o}gler}, {Shelyag}, {Sch{\"u}ssler},
  {Cattaneo}, {Emonet}, \& {Linde}}]{Vogler2005}
{V{\"o}gler}, A., {Shelyag}, S., {Sch{\"u}ssler}, M., {et~al.} 2005, \aap, 429,
  335

\bibitem[{{Voronov} {et~al.}(2022){Voronov}, {Yakovleva}, \&
  {Belyaev}}]{VoronovYakovleva2022}
{Voronov}, Y.~V., {Yakovleva}, S.~A., \& {Belyaev}, A.~K. 2022, \apj, 926, 173

\bibitem[{{{\v{S}}t{\v{e}}p{\'a}n} \& {Trujillo
  Bueno}(2013)}]{StepanTrujilloBueno2013}
{{\v{S}}t{\v{e}}p{\'a}n}, J. \& {Trujillo Bueno}, J. 2013, \aap, 557, A143

\bibitem[{{Watson} {et~al.}(2019){Watson}, {Hansen}, {Selsing}, {Koch},
  {Malesani}, {Andersen}, {Fynbo}, {Arcones}, {Bauswein}, {Covino}, {Grado},
  {Heintz}, {Hunt}, {Kouveliotou}, {Leloudas}, {Levan}, {Mazzali}, \&
  {Pian}}]{WatsonHansen2019}
{Watson}, D., {Hansen}, C.~J., {Selsing}, J., {et~al.} 2019, \nat, 574, 497

\bibitem[{{Wheeler} {et~al.}(2023){Wheeler}, {Abruzzo}, {Casey}, \&
  {Ness}}]{WheelerAbruzzo2023}
{Wheeler}, A.~J., {Abruzzo}, M.~W., {Casey}, A.~R., \& {Ness}, M.~K. 2023, \aj,
  165, 68

\bibitem[{{Yakovleva} {et~al.}(2020){Yakovleva}, {Belyaev}, \&
  {Bergemann}}]{YakovlevaBelyaev2020}
{Yakovleva}, S.~A., {Belyaev}, A.~K., \& {Bergemann}, M. 2020, Atoms, 8, 34

\end{thebibliography}

\begin{appendix}
\nolinenumbers

\section{The \Dispatch framework}
The \Dispatch high-performance simulation framework \citep{NordlundRamsey2018} utilises a task-based parallelisation scheme designed for 3D astrophysical applications. This framework organises computations, such as hydrodynamics or radiative transfer solvers, into discrete tasks.

To maintain the correct execution order, \Dispatch defines neighbour relations between these tasks. Two tasks are neighbours if the execution of one depends on the status of the other. A task becomes ready for execution once all its neighbours are ahead in time, indicating they have either finished or are scheduled for a later execution. The framework updates the time attribute of each task by adding a specific step after every execution. In this study, where the post-processing involves single frozen hydrodynamic snapshots or 1D hydrostatic atmospheres, the time attribute serves as a proxy for the number of lambda iterations.

\subsection{Atmosphere interpolation}

\begin{figure}[!h]
    \centering
    \includegraphics[scale=\pltscale]{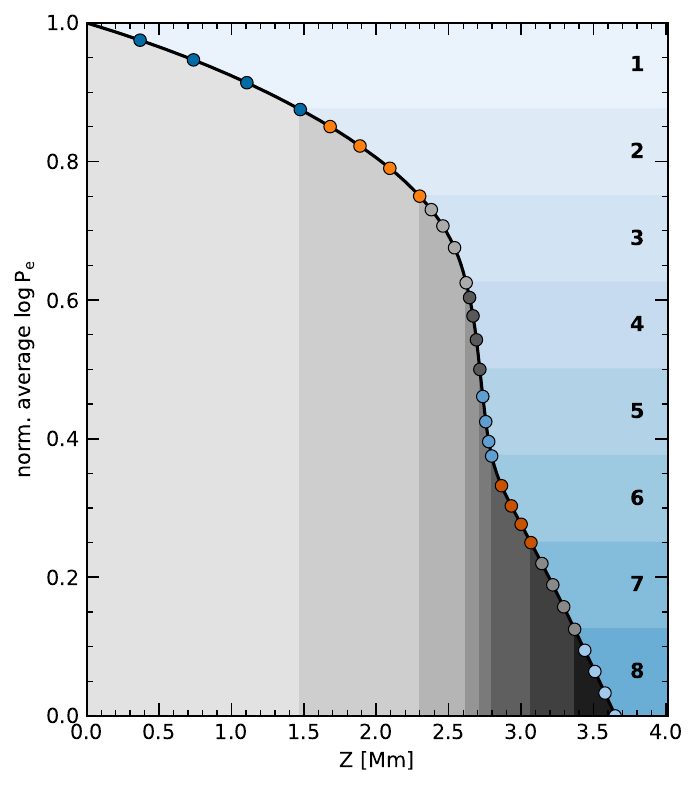}
    \caption{Visualisation of the atmosphere interpolation routine. Each atmosphere is split in regular intervals of the horizontally averaged logarithmic electron pressure. The resulting intervals are considered as individual subregions. Each subregion finally features regular equidistant meshes.}
    \label{fig:AtmosInterp}
\end{figure}

\Dispatch is tailored for regular grids within individual subregions, but it natively handles subregions of non-uniform sizes due to its focus on adaptive mesh-refinement. To ensure sufficient sampling of photospheric layers and to homogenise various input atmospheres while meeting \Dispatch grid requirements, \Code processes all input atmospheres as follows:

\begin{enumerate}
    \item For atmospheres on an optical depth scale, $\chi_{500}$ is determined at each point using the EOS and opacity package to convert the optical depth to a height scale.
    \item The logarithmic electron pressure, $\log(P_e)$, is determined using the EOS.
    \item For 3D atmospheres, the horizontal average of $\log(P_e)$ is computed.
    \item The resulting 1D atmosphere is divided into $N_{\rm sub}$ equal intervals of $\log(P_e)$.
    \item Height intervals corresponding to these $\log(P_e)$ intervals are determined.
    \item These height intervals serve as subregion boundaries. A regular grid with a resolution of $n_x \times n_y \times n_z/N_{\rm sub}$ is created for each subregion, where $n_x, n_y, n_z$ are specified as input (with $n_x = n_y = 1$ for 1D atmospheres). If $n_z/N_{\rm sub}$ is not a natural number, one or more subregions have one extra xy-plane.
    \item The original atmosphere is interpolated onto the new subregion grids using piecewise cubic monotonic Hermite interpolation.
\end{enumerate}

Figure~\ref{fig:AtmosInterp} illustrates steps 4--6, where equal intervals in $\log(P_e)$ are represented by blue shaded areas on the y-axis, and the resulting non-uniform height intervals are shown as grey shaded areas on the x-axis. Logarithmic electron pressure is used as a reference because it yields low resolution below the photosphere, high resolution near $\tau_{500}=1$, and intermediate resolution in regions of low optical depth.

\subsection{Organising NLTE radiative transfer in tasks}
\begin{figure}[!h]
    \centering
    \includegraphics[width=\linewidth]{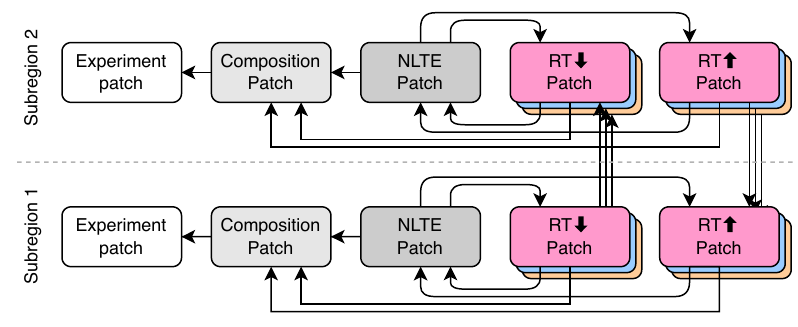}
    \caption{\Dispatch patch dependencies. Arrows point towards needed patches. Circular dependencies give rise to iterations.}
    \label{fig:TumultFlow}
\end{figure}

Consistent with other 3D NLTE spectrum synthesis codes, the stellar atmosphere is divided into multiple spatial subregions. During opacity calculations and radiative transfer, the frequency domain is further partitioned into multiple intervals. Figure~\ref{fig:TumultFlow} illustrates the organisation of the NLTE spectrum synthesis problem into patches and their respective relationships. Within the \Dispatch context, a patch is a task associated with a specific subregion.

\begin{itemize}
    \item Experiment patch: This patch stores the state variables and velocities for a single subregion of the model atmosphere. Executed once at the start, it crops and interpolates data from the original model atmosphere onto a regular grid covering the subregion.\\

    \item Composition patch: This patch handles microphysics and calculates partial pressures for all ions and molecules within the subregion defined by the Experiment patch. It also computes several frequency-independent intermediate quantities used during opacity calculations. This patch is executed once after state variables are loaded into the Experiment patch. \\

    \item NLTE patch: This optional patch is required when employing or computing NLTE level populations. During the first update, it either reads precomputed departure coefficients or initialises LTE populations for trace particles (atoms or molecules) specified in the input file. The patch subsequently solves the statistical equilibrium equations and updates level populations once radiative rates are determined for all frequency intervals in the subregion. \\

    \item RT patch: Two RT patches exist for each frequency interval per subregion. The UP type propagates radiation from the bottom to the top of the subregion, while the DOWN type propagates it from top to bottom. The latter is only initialised if NLTE departure coefficients are computed or if scattering is enabled. Each patch loops over the angles in a chosen quadrature and performs the following steps: \\
    \begin{enumerate}
        \item Calculating continuum opacities from the opacity package.
        \item Adding line opacities from the opacity table.
        \item Adding line opacities from the line list.
        \item Adding continuum opacities from trace particles.
        \item Adding line opacities from trace particles.
        \item Solving the radiative transfer from one end of the subregion to the other.
        \item Calculating radiative transition rates for trace particles. \\
    \end{enumerate}
    Steps 1 and 6 are always executed, while others are included only if the necessary input data is provided. RT patches require the prior execution of the Experiment, Composition, and optionally NLTE patches. They also connect different subregions, depending on RT patches for the same frequency interval located in the upstream direction (the UP patch below for UP patches, or the DOWN patch above for DOWN patches).
\end{itemize}

For completeness, initialising RT patches is optional; the code may also be used solely to read and process input atmosphere models, as described in the following section.

\subsection{Parallelisation}
\Code currently only uses the OpenMP (Open Multi-Processing) capabilities of \Dispatch, which also supports MPI (Message Passing Interface) parallelisation across multiple nodes. Due to the reduced accuracy losses from downsampling (see Sect.~\ref{sec:projection}), the memory requirements are much reduced and even our largest NLTE models, for example Fe, run comfortably on a single node (see Table~\ref{tab:benchmarks_comparison}). While OpenMP allows for a shared memory layout, the code has been specifically designed to have a self-contained memory layout for each interval in the spatial-frequency domain, in case of switching to MPI later on. Global communication is only necessary for the convergence criterion. Arrays of data, needed for the radiative transfer, are only "downloaded" from the upstream neighbours of the RT patches. The only exception occurs if scattering is enabled. The mean intensity has to be stored for each spatial-frequency point. The user can decide whether to store this data on disk between iterations or keep it in memory.

\clearpage
\section{Optimised computational framework for line opacity} \label{sec:line_kernel}

The calculation of the monochromatic linear absorption coefficient $\chi_{\nu}$ and emissivity $\eta_{\nu}$ represents the primary computational burden in 3D NLTE spectrum synthesis with many lines. For a transition from a lower level $i$ to an upper level $j$ and under the approximation of complete redistribution, these are defined as:
\begin{align}
    \chi_{\nu} &= \frac{h\nu}{4\pi} (n_i B_{ij} - n_j B_{ji}) \phi_{\nu} \\
    \eta_{\nu} &= \frac{h\nu}{4\pi} n_j A_{ji} \phi_{\nu}
\end{align}
To achieve high performance, we transform these expressions to isolate frequency-independent terms. We define a dimensionless intensity $\mathcal{I}_{\nu} = I_{\nu} / \mathcal{S}_{\nu}$, where the scaling factor $\mathcal{S}_{\nu} = 2h\nu^3/c^2$ corresponds to the photo-emissive capacity. Under this scaling, the emissivity transforms to:
\begin{equation}
    \eta^*_{\nu} = \frac{\eta_{\nu}}{\mathcal{S}_{\nu}} = \frac{h\nu}{4\pi} n_j B_{ji} \phi_{\nu}
\end{equation}
where we have utilised the Einstein relation $A_{ji} = \frac{2h\nu^3}{c^2} B_{ji}$.
The factor $\mathcal{S}_{\nu}$ is eventually multiplied to the final output spectrum.
Incorporating the departure coefficients $b_i = n_i / n_i^{\text{LTE}}$ and the Boltzmann factor $\beta_{\nu} = \exp(-h\nu/kT)$, we may reformulate

\begin{align}
    \chi_{\nu} &= \frac{h\nu}{4\pi} \phi_{\nu} n_i^{\text{LTE}} B_{ij}\,(b_i - \beta_\nu\,b_j ) \\
    \eta^*_{\nu} &= \frac{h\nu}{4\pi} \phi_{\nu} n_i^{\text{LTE}} B_{ij}\,\beta_\nu\,b_j,
\end{align}

where we have made use of the relation of the LTE populations

\begin{equation}
    \left[\frac{n_j}{n_i}\right]^{\text{LTE}} = \frac{g_j}{g_i} \beta_{\nu} = \frac{B_{ij}}{B_{ji}} \beta_{\nu}.
\end{equation}

\subsection{Separation of variables}
The normalised Voigt profile $\phi_{\nu}$ is typically expressed as:
\begin{equation}
    \phi_{\nu} = \frac{1}{\sqrt{\pi} \Delta\nu_D} H(a, v)
\end{equation}
where the Doppler width $\Delta\nu_D$ and the dimensionless frequency $v = (\nu - \nu_0)/\Delta\nu_D$ vary spatially. We introduce the dimensionless inverse Doppler width, $\mathcal{W}_D = \nu_0 / \Delta\nu_D$. This is very convenient as $\mathcal{W}_D$ is only isotope dependent, hence can be computed once in the outermost loop. By defining scaled Einstein coefficients $B^*_{ij} = \frac{h}{4\pi} B_{ij}$, we define a line-specific opacity kernel $\mathcal{K}_{ij}$ at each spatial point:
\begin{equation}
    \mathcal{K}_{ij} = n_i^{\text{LTE}} B^*_{ij} \mathcal{W}_D
\end{equation}
This kernel represents the integrated opacity of the line in LTE, scaled by the Doppler factor. Here we have replaced $\nu$ by $\nu_0$, which is a perfectly reasonable approximation for the small wavelength windows covered by each line. Then we define the pre-factors:
\begin{align}
    \mathcal{U}_{ij} &= \mathcal{K}_{ij} b_i \\
    \mathcal{U}_{ji} &= \mathcal{K}_{ij} b_j
\end{align}

\subsection{Inner loop minimisation}
The innermost loop over frequency $\nu$ now computes the reduced variables $\hat{\chi}$ and $\hat{\eta}$:
\begin{align}
    \hat{\chi}_{\nu} &= \sum_{\text{isotopes}} \sum_{\text{lines}} \mathcal{U}_{ij} H^*(a, v) \\
    \hat{\eta}_{\nu} &= \sum_{\text{isotopes}} \sum_{\text{lines}} \mathcal{U}_{ji} H^*(a, v)
\end{align}
where $H^*(a, v) = H(a, v)/\sqrt{\pi}$. The physical coefficients are recovered via a final frequency-dependent correction to account for stimulated emission:
\begin{align}
    \eta^*_{\nu} &= \hat{\eta}_{\nu} \beta_\nu \\
    \chi_{\nu} &= \hat{\chi}_{\nu} - \eta^*_{\nu}
\end{align}
This formulation ensures that the transcendental Voigt function is only multiplied by two pre-calculated scalars per frequency point, and the frequency-dependent exponential $\beta_\nu$ is applied only once to the sum of all lines, rather than per line.

\clearpage
\section{Code validation} \label{sec:validation}

\subsection{NLTE comparison with \Multi} \label{sec:validation_1d}

\begin{table}[!h]
\centering
\small
\setlength{\tabcolsep}{4pt}
\caption{Atomic data for the Mg, Fe and O lines used in the model atoms for the code validation.}
\label{tab:nlte_lines}
\begin{tabular}{lccccc}
\toprule
species & $\lambda_0$ [\AA] & E$_i$ [eV] & $\log_{10}(gf)$ & $\gamma_{\rm vdw}$ & $\gamma_{\rm rad}$ \\
\midrule
Mg I  & 4571.10 & 0.00 & -5.397 & 222.253  & 6.70e+02 \\
Mg I  & 4702.99 & 4.35 & -0.456 & 2806.269 & 5.01e+08 \\
Mg I  & 5172.68 & 2.71 & -0.363 & 729.238  & 9.77e+07 \\
Mg I  & 5183.60 & 2.71 & -0.168 & 729.238  & 9.77e+07 \\
Mg I  & 5528.42 & 4.35 & -0.547 & 1461.310 & 5.25e+08 \\
\addlinespace
Fe I  & 4235.94 & 2.43 & -0.247 & 739.240 & 1.51e+08  \\
Fe I  & 4250.79 & 1.56 & -0.591 & 310.262 & 9.12e+07  \\
Fe I  & 5269.54 & 0.86 & -1.060 & 237.249 & 1.51e+07  \\
\addlinespace
O I   & 6300.30 & 0.00 & -9.720 & 1.0     & 2.110e-05  \\
O I   & 7771.94 & 9.15 & 0.369  & 453.234 & 3.890e+07  \\
O I   & 7775.39 & 9.15 & 0.002  & 453.234 & 3.890e+07  \\
\bottomrule
\end{tabular}
\tablefoot{Van der Waals damping coefficients $\gamma_{\rm vdw}$ of order unity refer to scaling factors of the \citet{Unsold1955} recipe, while the other values correspond to the recipe from \citet{BarklemPiskunov2000}.}
\end{table}
 
\begin{figure*}
    \centering
    \includegraphics[scale=\pltscale, trim={0 30 0 0}, clip]{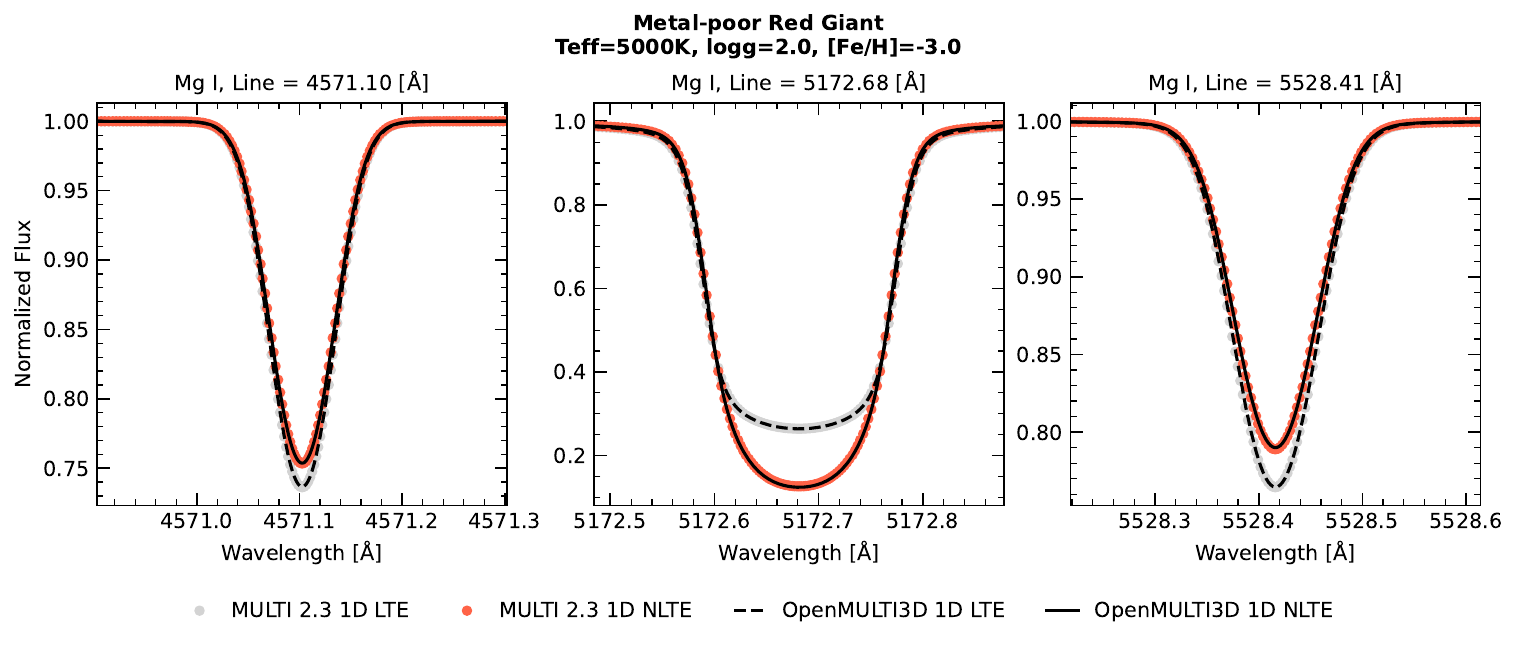}
    \includegraphics[scale=\pltscale]{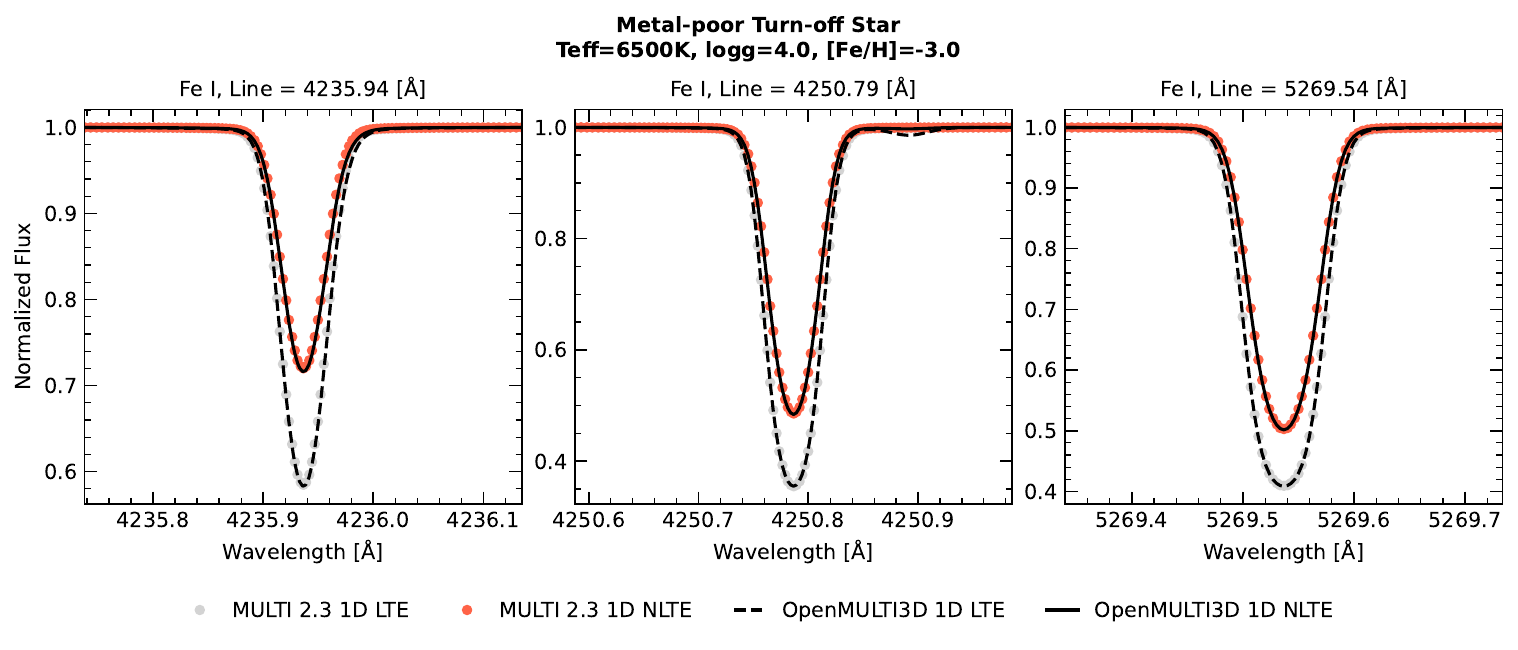}
    \caption{NLTE Mg and Fe line profile test calculations using 1D \Marcs atmospheres comparing \Multi (version 2.3) and \Code with matching settings. The extra line blend in the right wing of the bottom centre panel does not show up in the MULTI 2.3 results, as it does not take into account overlapping lines.}
    \label{fig:nlte_validation}
\end{figure*}

We compare NLTE effects predicted by \Multi (version 2.3) and \Code for three \spc{Mg}{1} lines in a metal-poor red giant (\stellarparams{5000}{2.0}{-3}) and three \spc{Fe}{1} lines in a metal-poor turn-off star (\stellarparams{6500}{4.0}{-3}). The resulting 1D LTE and 1D NLTE line profiles are presented in Fig.~\ref{fig:nlte_validation}. We selected these specific elements as the \spc{Mg}{1} model from \citet{BergemannCollet2017}, containing $86$ energy levels and $327$ transitions, corresponds to one of our smaller NLTE models, while the Fe \textsc{\tiny I\,\&\,II} model from \citet{BergemannLind2012}, having $607$ energy levels and $13\,241$ transitions, is our largest atomic NLTE model. Our line and atmosphere selection is based on previous high-resolution studies of abundances of Galactic stars \citep{BergemannLind2012, BergemannCollet2017}. The three \spc{Fe}{1} lines were chosen as they are observable in spectra of metal-poor stars of turn-off, RGB and red supergiant type \citep[e.g. Table~A1 in][]{BergemannLind2012}. The same holds for the selected \spc{Mg}{1} lines \citep[e.g. Table~5 in][]{BergemannCollet2017}. The Mg lines are shown for a representative metal-poor RGB  model, while the Fe lines are displayed for a metal-poor turn-off star. The \Code configuration was carefully matched to the \Multi setup, meaning that outdated atomic and molecular data have been used in this comparison.

In all cases, the agreement between the two codes is excellent. Any residual differences are negligible and remain well within the expected numerical precision of the radiative transfer calculations. These results demonstrate that \Code reproduces the NLTE line formation predicted by \Multi with high fidelity across different stellar parameters and atomic species.

\subsection{3D NLTE comparison with \MultiD} \label{sec:validation_3d}

\begin{figure*}
    \centering
    \includegraphics[scale=\pltscale, trim={0 30 0 0}, clip]{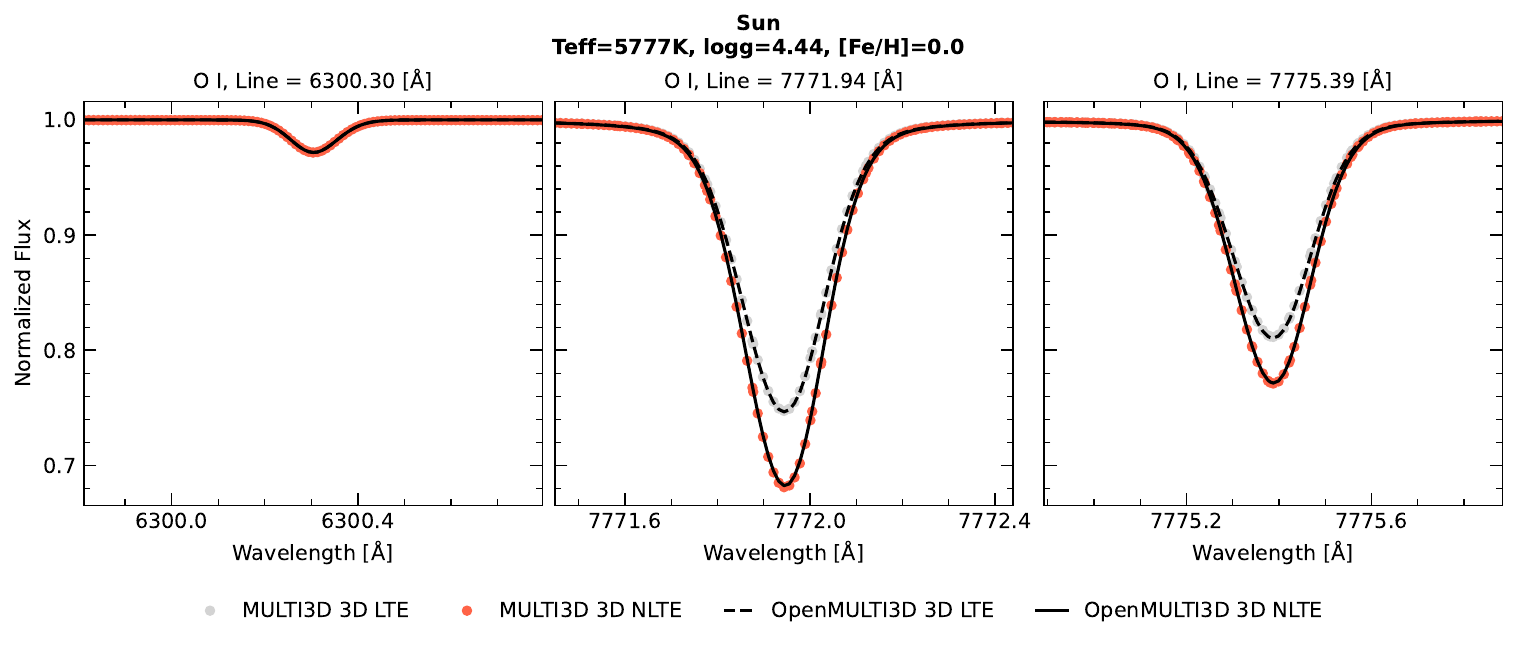}
    \includegraphics[scale=\pltscale]{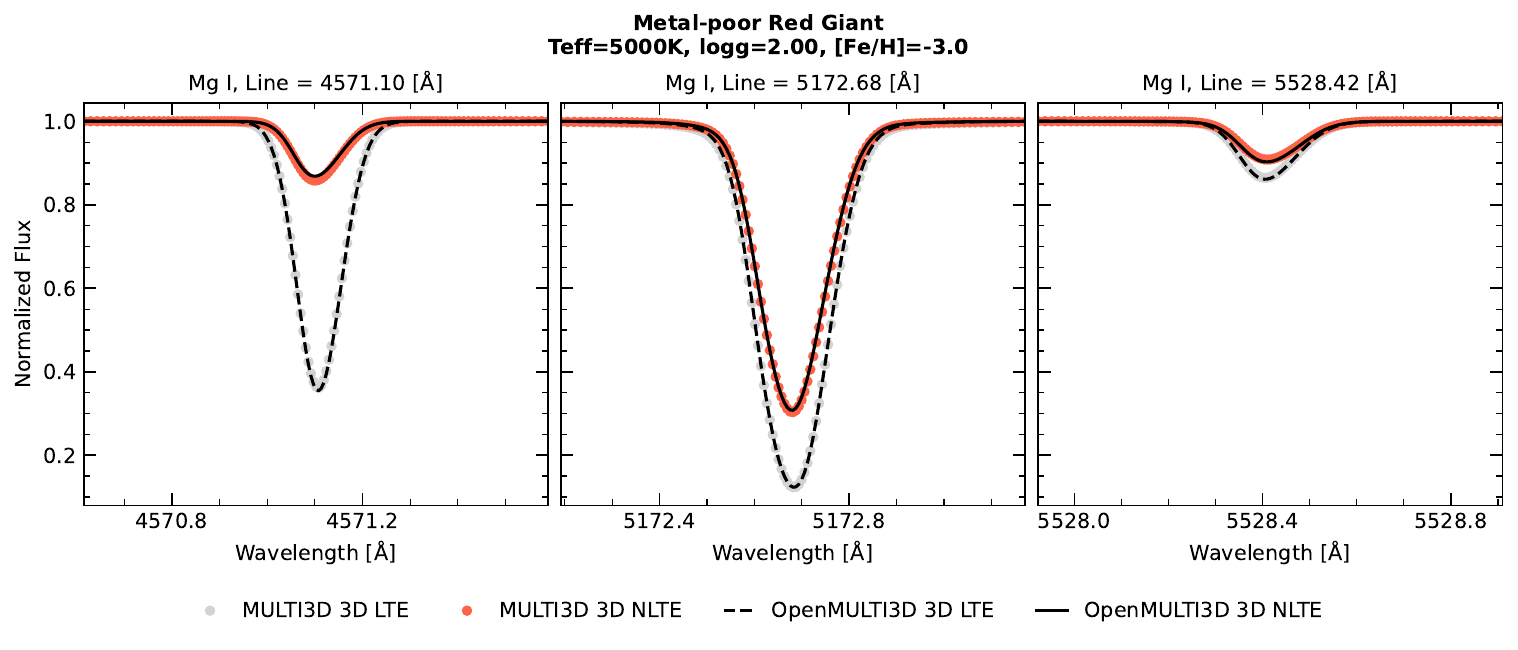}
    \caption{3D NLTE O and Mg line profile test calculations using 3D \Stagger atmospheres comparing \MultiD and \Code with matching settings.}
    \label{fig:3dnlte_validation}
\end{figure*}

During our code validation we have identified three issues with the original \MultiD code.

\begin{enumerate}
    \item When downsampling the full-resolution snapshots, the implemented interpolation routine is fixed to match the first and last point along each dimension. This is counterproductive in the case of periodic boundary conditions in the horizontal as downsampling from 240 to 120 points along the x or y directions does not simply probe every 2nd point for regularly spaced grids. Instead the new spacing is slightly larger than double the original spacing as it always includes the last point. This introduces artefacts at the horizontal boundaries as the first and last point are incorrectly assumed to have the same spacing as the other points after downsampling. The correct approach is to fix the first point and determine the position of the last point from the new spacing which is consistent across the boundary.
    \item The temperature, mass density and electron density are interpolated linearly onto the downsampled grid. This tends to overestimate the interpolated densities and defeats the purpose of using a non-linear formal radiative transfer solver. Instead we propose to use higher-order interpolation in logarithmic quantities as is done during the hydrodynamic simulations themselves \citep[e.g.][]{SteinNordlund2024}.
    \item The horizontal velocities are incorrectly flipped when being projected onto the direction of the photon path for outgoing rays. This results in the interpolations and the projection of the velocities  to the photon path using inconsistent angles with respect to one another. This inconsistency, however, only produces subtle errors in the flux line profiles, therefore it most likely did not affect previous abundance calculations noticeably.
\end{enumerate}

Figure~\ref{fig:3dnlte_validation} presents the comparison of 3D LTE and 3D NLTE line profiles computed with \Code and \MultiD, after fixing 1 and 3 of the aforementioned issues, for three \spc{Mg}{1} lines and three \spc{O}{1} lines. For the \spc{Mg}{1} lines we reuse the same stellar parameters for a typical metal-poor red giant as before while for the \spc{O}{1} line we have chosen a solar atmosphere model. In both cases we employ a single \Stagger snapshot as the \Stagger format is natively supported by both codes.

The presented spectral lines have been computed with a downsampled snapshot resolution of $80\times80\times230$ spatial points. This is the recommended resolution by the \Stagger community as it provides an optimal trade-off between computational efficiency and precision in the normalised line flux \citep{RodriguezDiazLagae2024}, assuming one does not pre-tilt the atmospheres. We further use the Carlsson A4 angle quadrature (12 rays per hemisphere) for the short-characteristics solver, which is the default setting in \MultiD, and choose a Lobatto quadrature with 4 inclinations and 4 azimuthal directions (13 rays per hemisphere) for the long-characteristics solver which is a common choice in 3D NLTE spectrum synthesis \citep{BergemannHoppe2026}. The results from each code agree well across both stellar types. As in Sect.~\ref{sec:validation_1d}, this test has been carried out by matching the \Code inputs to the \MultiD inputs, meaning that outdated atomic and molecular data have been used.

\subsection{SED comparison with \Turb} \label{sec:turbo_validation}

\begin{figure}
    \centering
    \includegraphics[scale=\pltscale]{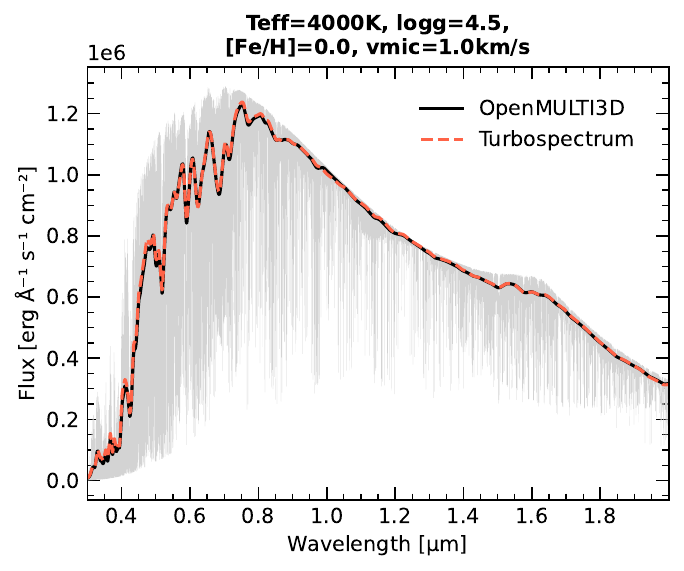}
    \includegraphics[scale=\pltscale]{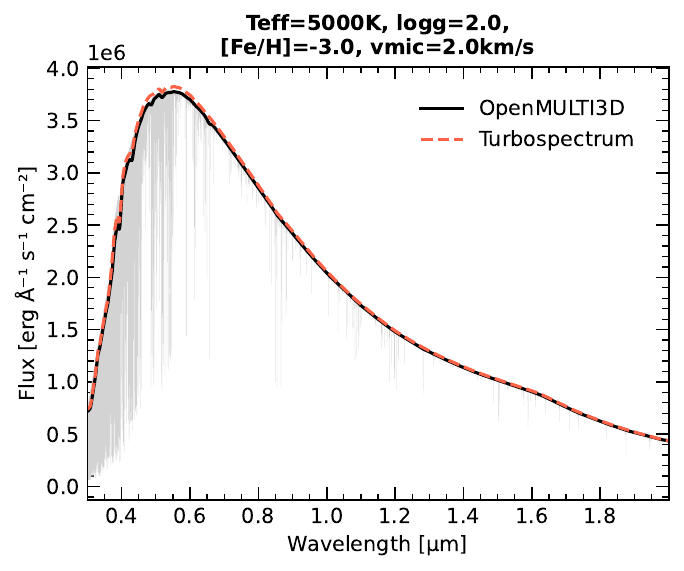}
    \includegraphics[scale=\pltscale]{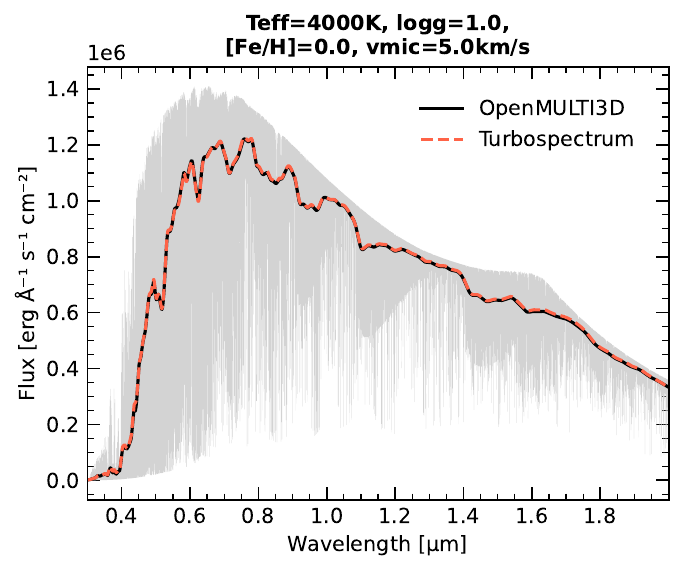}
    \caption{Line list test calculation of a typical M dwarf, metal-poor red giant and red supergiant computed in 1D LTE using \Code and \Turb. Grey lines in the background show the raw fluxes from \Code, while the solid black shows the same flux, but convolved to a resolution of $R = 100$. The dashed red line shows the analogous \Turb result.}
    \label{fig:sed_validation}
\end{figure}

Figure~\ref{fig:sed_validation} compares the 1D LTE spectral energy distributions computed with \Code and \Turb over the wavelength range \SIrange{3000}{20000}{\angstrom}. The comparison is shown for the stellar model atmospheres from Table~\ref{tab:stellar_parameters}, including a typical M dwarf, a metal-poor red giant, and a red supergiant. The spectra are overplotted after convolution to a resolving power of $R = 100$ to enable an easier visual comparison between both codes.

The 1D LTE SEDs computed with \Turb and \Code are in excellent agreement across the full spectral range. The overall continuum shapes, as well as the detailed structure of molecular and atomic features, are reproduced consistently by both codes. The level of agreement confirms the consistency of the radiative transfer implementation, line list, and continuous opacity treatment in \Code relative to \Turb.

\begin{table}
\centering
\caption{Optical benchmark lines used in the projection scheme validation.}
\label{tab:benchmark_lines}
\begin{tabularx}{\linewidth}{
>{\centering\arraybackslash}X
>{\centering\arraybackslash}X
>{\centering\arraybackslash}X
>{\centering\arraybackslash}X}
\toprule
species & $\lambda_0$ [\AA] & E$_i$ [eV] & $\log(gf)$ \\
\midrule
CH & 4218.71 & 0.413 & -1.315 \\
   & 4218.73 & 0.413 & -1.337 \\
   & 4218.74 & 1.484 & -4.377 \\
   & 4218.78 & 0.413 & -3.410 \\
\noalign{\vskip 2mm}
Ca I & 6162.17 & 1.899 & -0.170 \\
\noalign{\vskip 2mm}
Fe I & 6151.62 & 2.176 & -3.295 \\
     & 6151.69 & 5.012 & -3.761 \\
\noalign{\vskip 2mm}
Fe II & 4923.92 & 2.891 & -1.260 \\
\noalign{\vskip 2mm}
Mg I & 5528.40 & 4.346 & -0.547 \\
\noalign{\vskip 2mm}
Na I & 5889.95 & 0.000 & 0.108 \\
\noalign{\vskip 2mm}
Ni I & 6767.77 & 1.826 & -2.140 \\
\noalign{\vskip 2mm}
O I & 7774.17 & 9.146 & 0.203 \\
\noalign{\vskip 2mm}
Ti II & 5336.79 & 1.582 & -1.600 \\
\bottomrule
\end{tabularx}
\end{table}
 
\clearpage
\section{Line damping recipes} \label{sec:damping}

Currently supported broadening mechanisms cover radiative broadening, Stark broadening and pressure broadening from collisions with neutral hydrogen (van der Waals broadening). The normalised Voigt profile $\phi_{\nu}$ is typically expressed as:
\begin{equation}
    \phi_{\nu} = \frac{1}{\sqrt{\pi} \Delta\nu_D} H(a, v),
\end{equation}

where H is the Voigt function and Doppler width $\Delta\nu_D$ and damping parameter $a$ are given by

\begin{align}
    \Delta\nu_D &= \dfrac{\nu_0}{c}
\sqrt{\dfrac{2kT}{m} + v_{\rm mic}^2} \\
    a &= \frac{1}{4 \pi \Delta\nu_D}\left(\Gamma_{\rm radiative} + \Gamma_{\rm vdw} + \Gamma_{\rm stark}\right). \label{eq:damping}
\end{align}

The widths $\Gamma_{\rm radiative}$, $\Gamma_{\rm vdw}$ and $\Gamma_{\rm stark}$ in Eq.~(\ref{eq:damping}) are full widths at half maximum in angular frequency (\si{\radian\per\second}). They are evaluated in every grid cell from the local temperature $T$, electron number density $N_e$ and number density of neutral hydrogen $N_{\rm \spc{H}{1}}$. The van der Waals broadening is due to collisions with neutral hydrogen atoms and Stark broadening is due to electron impacts. Each line list record carries three broadening entries, $\gamma_{\rm rad}$, $\gamma_{\rm vdw}$ and $\gamma_{\rm stk}$ (Table~\ref{tab:llist_demo}). The value of each entry selects the recipe in Table~\ref{tab:damping} that converts it into the corresponding $\Gamma$, so that line lists from different sources can be combined without conversion. The same recipes apply to the bound-bound transitions of the model atoms.

For $\gamma_{\rm rad}=0$ we adopt the estimate of \citet{WheelerAbruzzo2023}, $g_j A_{ji}$, which requires only $g_i f_{ij}$. \Turb divides it by the statistical weight $g_j$ of the upper level, giving $A_{ji}$. Both forms are rough lower limits, since the radiative width is the sum of the radiative decay rates out of both levels, and most line lists do provide $\gamma_{\rm rad}$. As this estimate would be the only use of $g_j$ in the line opacity calculation, we have decided to adopt the \citet{WheelerAbruzzo2023} approach and kept $g_j$ out of the line list format.

\begin{table*}
  \caption{Damping recipes for metal and molecular lines.}
  \label{tab:damping}
  \centering
  \renewcommand{\arraystretch}{1.4}
  \begin{tabularx}{\textwidth}{@{}r@{}l >{\raggedright\arraybackslash}X l >{\raggedright\arraybackslash}X@{}}
    \toprule
    \multicolumn{2}{@{}l}{Entry} & Meaning of the entry & Width & Reference \\
    \midrule
    \multicolumn{5}{@{}l}{Radiative broadening} \\
    & $\gamma_{\rm rad} > 0$ & radiative damping rate $\Gamma_{\rm radiative}$ in \si{\radian\per\second} & $\gamma_{\rm rad}$ & \\
    & $\gamma_{\rm rad} = 0$ & classical estimate & $8\pi^2 e^2 \nu_0^2\, g_i f_{ij} / (m_e c^3) = g_j A_{ji}$ & \citet{WheelerAbruzzo2023} \\
    \midrule
    \multicolumn{5}{@{}l}{van der Waals broadening} \\
    & $\gamma_{\rm vdw} \ge 20$ & packed $\sigma.\alpha$\tablefootmark{a,b} & $N_{\rm \spc{H}{1}}\, 2\, K_\alpha\, v_0\, \sigma\, (\bar v/v_0)^{1-\alpha}$ & \citet{BarklemPiskunov2000} \\
    $0 <{}$ & $\gamma_{\rm vdw} < 20$ & Uns\"old enhancement factor\tablefootmark{a,c}, dimensionless & $N_{\rm \spc{H}{1}}\, \gamma_{\rm vdw}\, 17\, C_6^{2/5}\, \bar v^{3/5}$ & \citet{Unsold1955} \\
    $-20 \le{}$ & $\gamma_{\rm vdw} < 0$ & $\log_{10}(\Gamma_{\rm vdw}/N_{\rm \spc{H}{1}})$ at \SI{e4}{K}\tablefootmark{f}, with $\Gamma_{\rm vdw}/N_{\rm \spc{H}{1}}$ in \si{\radian\centi\metre\cubed\per\second} & $N_{\rm \spc{H}{1}}\, 10^{\gamma_{\rm vdw}}\, (T/\SI{e4}{K})^{0.3}$ & \citet{Unsold1955} \\
    & $\gamma_{\rm vdw} < -20$ & $\log_{10} C_6$\tablefootmark{a}, with $C_6$ in \si{\centi\metre\tothe{6}\per\second} & $N_{\rm \spc{H}{1}}\, 17\, C_6^{2/5}\, \bar v^{3/5}$ & \citet{Unsold1955}, \citet{Gray2005} \\
    & $\gamma_{\rm vdw} = 0$ & -- & $0$ & \\
    \midrule
    \multicolumn{5}{@{}l}{Quadratic Stark broadening} \\
    & $\gamma_{\rm stk} > 0$ & $\Gamma_{\rm stark}/N_e$ in \si{\radian\centi\metre\cubed\per\second} & $N_e\, \gamma_{\rm stk}$ & \\
    $-10 \le{}$ & $\gamma_{\rm stk} < 0$ & $\log_{10}(\Gamma_{\rm stark}/N_e)$ at \SI{e4}{K}\tablefootmark{f}, with $\Gamma_{\rm stark}/N_e$ in \si{\radian\centi\metre\cubed\per\second} & $N_e\, 10^{\gamma_{\rm stk}}\, (T/\SI{e4}{K})^{1/6}$ & \citet{Unsold1955} \\
    & $\gamma_{\rm stk} < -10$ & $\log_{10} C_4$\tablefootmark{d}, with $C_4$ in \si{\centi\metre\tothe{4}\per\second} & $N_e\, 39\, v_e^{1/3}\, C_4^{2/3}$ & \citet{Unsold1955}, \citet{Gray2005} \\
    & $\gamma_{\rm stk} = 0$ & Cowley approximation\tablefootmark{e} & $N_e\, c_Z\, (n^*)^4$ & \citet{Cowley1971} \\
    \bottomrule
  \end{tabularx}
  \tablefoot{The recipe is selected by the value of the line list entries $\gamma_{\rm rad}$, $\gamma_{\rm vdw}$ and $\gamma_{\rm stk}$. Units are CGS and energies are in \si{eV}. All widths $\Gamma$ are full widths at half maximum in \si{\radian\per\second}, and the number densities $N_{\rm \spc{H}{1}}$ and $N_e$ are in \si{\per\centi\metre\cubed}. $E_i$ and $E_j$ are the excitation energies of the lower and upper level, $E_{\rm ion}$ is the ionisation energy of the absorbing species and $Z$ its ionisation stage ($Z=1$ for neutral atoms).
    \tablefoottext{a}{$\bar v = [8kT/\pi\,(1/m_{\rm H} + 1/m)]^{1/2}$ is the mean relative speed of a hydrogen atom and the absorber of mass $m$.}
    \tablefoottext{b}{The integer part is the cross-section $\sigma$ in units of $a_0^2$ and the fractional part the velocity exponent $\alpha$, with $K_\alpha = (4/\pi)^{\alpha/2}\, \Gamma(2-\alpha/2)$, where $\Gamma(x)$ denotes the gamma function, and $v_0 = \SI{e4}{\metre\per\second}$.}
    \tablefoottext{c}{$C_6 = C_{\rm U} Z^2 [(E_{\rm ion}-E_j)^{-2} - (E_{\rm ion}-E_i)^{-2}]$ with $C_{\rm U} = \SI{3.22e-31}{\centi\metre\tothe{6}\per\second\electronvolt\squared}$.}
    \tablefoottext{d}{$v_e = (8kT/\pi m_e)^{1/2}$ is the mean electron speed.}
    \tablefoottext{e}{$n^* = Z\,[E_{\rm H}/(E_{\rm ion}-E_j)]^{1/2}$ is the effective principal quantum number of the upper level, with $E_{\rm H} = \SI{13.6}{eV}$. The coefficient $c_Z = \SI{2.26e-7}{\radian\centi\metre\cubed\per\second}$ for $Z=1$ and $c_Z = \SI{5.42e-7}{\radian\centi\metre\cubed\per\second}\,(Z+1)^{-2}$ for $Z>1$ follows from \citet{Cowley1971} evaluated at \SI{e4}{K}, which he recommends using at all temperatures. $\Gamma_{\rm stark}=0$ for $E_i \ge E_{\rm ion}$.}
    \tablefoottext{f}{The temperature scaling follows from the impact widths $\Gamma_{\rm vdw} \propto \bar v^{3/5}$ and $\Gamma_{\rm stark} \propto v_e^{1/3}$ \citep{Unsold1955}.}
  }
\end{table*}
 
\end{appendix}

\end{document}